\documentclass[11pt, fleqn, a4paper]{article}
\usepackage[usenames]{color} 
\usepackage{amssymb} 
\usepackage{amsmath}
\usepackage[utf8]{inputenc} 
\usepackage{braket, tensor, bm}
\usepackage[dvipdfmx]{graphicx}
\usepackage{mediabb}
\usepackage{here}
\usepackage{mathptmx}
\usepackage{algorithm,algorithmic}
\usepackage{authblk}

\makeatletter
\@addtoreset{equation}{section}

\makeatother

\makeatletter
\@addtoreset{figure}{section}

\makeatother

\makeatletter
\@addtoreset{table}{section}

\makeatother

\title{A Parallel Computing Method for the Coupled-Cluster Singles and Doubles}
\author[1]{Takumi Yamashita}  \author[2]{Taichi Kosugi}  \author[2]{Yu-ichiro Matsushita}  \author[3]{Tetsuya Sakurai}
\affil[1]{{\small Faculty of Engineering, Information and Systems, University of Tsukuba, 1-1-1 Tennodai, Tsukuba, Ibaraki 305-8573 Japan}}
\affil[2]{{\small Laboratory for Materials and Structures, Institute of Innovative Research, Tokyo Institute of Technology, Yokohama, Kanagawa 226-8503, Japan}}
\affil[3]{{\small Department of Computer Science, University of Tsukuba, 1-1-1 Tennodai, Tsukuba, Ibaraki 305-8573 Japan}}
\date{}

\begin{document}
 
 \parindent = 10pt
 
 \maketitle
  \begin{abstract}
    In this paper, we present a parallel computing method for the coupled-cluster singles and doubles (CCSD) in periodic systems.
    The CCSD in periodic systems solves simultaneous equations for single-excitation and double-excitation amplitudes.
    In the simultaneous equations for double-excitation amplitudes, each equations are characterized by four spin orbitals and three independent momentums of electrons.
    One of key ideas of our method is to use process numbers in parallel computing to identify two indices which represent momentum of an electron.
    When momentum of an electron takes $N_ {\bm{k}}$ values, $N_ {\bm{k}}^2$ processes are prepared in our method.
    Such parallel distribution processing and way of distribution of known quantities in the simultaneous equations
    reduces orders of computational cost and required memory scales by $N_ {\bm{k}}^2$ compared with a sequential method.
    In implementation of out method, communication between processes in parallel computing appears in the outmost loop in a nested loop but does not appear inner nested loop.
  \end{abstract}

 \section{Introduction}  \label{Introduction}
 
  The development of high-precision first-principles calculation methodologies stays an important research subject for the theoretical condensed-matter physics.
  Reflecting the recent growth of computational power,
  the wavefunction theories are getting attention not only in the quantum chemistry communities but also in the physics ones
  because they provide typically systematic ways to improve the accuracy,
  compared to the development of exchange correlation functionals for DFT calculations\cite{HK64, KS65}.
  For example, the density-matrix-renormalization group (DMRG)\cite{White92} has been applied to isolated molecular systems such as H$_2$O \cite{WM99}.
  The transcorrelated method\cite{BH69} was reported to be applied to uniform electron gases\cite{UTOSC05} and semiconductors\cite{OAT17}.
  Self-energy functional theory\cite{Potthoff03, Potthoff14} has recently been applied to isolated transition metal atoms\cite{KNFM18}.
  The coupled-cluster singles-and-doubles (CCSD)\cite{HJO00} has been applied to electron gases\cite{MLWMRLBC16} and various periodic systems\cite{MSCB17}.
  Furthermore, the one-particle spectra from the Green's functions (GFs) in CCSD method (GFCCSD) have also been reported for realistic systems\cite{KNFM18,FKNM18,NKFM18,PK18}.
  The relation between the equation of motion (EOM)-CCSD, a part of the GFCCSD procedure,
  and the $GW$ method\cite{Hedin65}, which is already known as a high-precision calculation method in condensed-matter physics,
  has been examined in detail,
  suggesting the high potential of CCSD-based methods compared with the $GW$ method\cite{LB18}.

  However, while the CCSD method in periodic systems is expected as a high-precision calculation, its large computational cost and large required memory capacity are an obstacle for its practical applications.
  Actually, the computational cost is
  $N_{\rm band}^6 N_{\bm{k}}^{4}$, with $N_{\bm{k}}$ being the number of $k$ points and $N_{\rm band}$ the number of bands.
  For the required memory regions, the bottlenecks are the antisymmetrized two-electron integrals and double-excitation amplitudes:
  Required memory space scales as $N_{\rm band}^4N_k^3$.
  In this context, an interpolation method has been developed to reduce the computational costs.
  In fact, a Wannier interpolation technique which interpolates $k$ points for self-energy has been developed \cite{KM19}.
  On the other hand, a straightforward approach to the problem of the required large memory space is large-scale memory distribution using supercomputers.
  Actually, the usage of supercomputers is effective for shortening the calculation time as well.
  Thus, the algorithm development of minimizing the communication time in accordance with large-scale memory distribution is a serious problem.

  In this study, we developed a new method for the CCSD method in periodic systems.
  It enables efficient memory distribution in large-scale parallel computing suppressing the communication time compared with a na\"ive parallel computing implementation.
  Actually, required memory space in each process can be suppressed to $N_{\rm band}^4 N_{\bm{k}}$ by using this new method.
  Note that this new method uses key ideas and some techniques for a parallel computing method \cite{YS19} for the Higher Order Tensor Renormalization Group (HOTRG) \cite{XCQZYX12}.
  This method for the HOTRG has successfully been applied \cite{AKYY19} to investigation of phase transition in the four-dimensional Ising model.
  
  This paper is organized as follows.
  In Section \ref{BasicEqs}, basic equation as a start point of the presented method is explained.
  In Section \ref{OutlineBasicEq}, an outline of the presented method is described and rearranged simultaneous equations for the presented method are given.
  In Section \ref{ImplementMethod}, implementation of the presented method is described.
  Section \ref{Conclusion} is devoted to conclusion.

 \section{Basic equations in the CCSD in a periodic system}  \label{BasicEqs}
 
 In the CCSD in a periodic system, basic equations are simultaneous equations.
 Unknown quantities are as follows:
  \begin{itemize}
      \item single-excitation amplitudes $t_{p \bar{\bm{k}}}^{g \bar{\bm{k}}}$
      \item double-excitation amplitudes $t_{i \bm{k}_i j \bm{k}_j}^{a \bm{k}_a b \bm{k}_b}$.
  \end{itemize}
 Known quantities are as follows:
  \begin{itemize}
      \item the matrix elements of the Fock operator $f_{p \bar{\bm{k}}}^{g \bar{\bm{k}}}$, $f_{p \bar{\bm{k}} q \bar{\bm{k}}}$ and $f^{g \bar{\bm{k}} h \bar{\bm{k}}}$
      \item antisymmetrized two-electron integrals $\bra{w \bm{k}_w x \bm{k}_x} \ket{y \bm{k}_y z \bm{k}_z}_{\phi_w \phi_x \phi_y \phi_z}$
  \end{itemize}
  A lower case letter represents a spin orbital.
  If it is a subscript (superscript), then the spin orbital is occupied (virtual).
  For antisymmetrized two-electron integrals, if a symbol $\phi_u$ $( u: w, x, y, z )$ is ``$o$'' (``$v$''), then the spin orbital $u$ is occupied (virtual).
  A symbol $\bm{k}$ with an index represents momentum of an electron in the spin orbital which is represented by the index.
  Single-excitation amplitudes and the matrix elements of the Fock operator consider two spin orbitals.
  They may have a nonzero value only when momentums of an electron in the two spin orbitals have the same value because of conservation law of momentum.
  Then, a symbol $\bar{\bm{k}}$ represents the common momentum of an electron in these two spin orbitals.

 Introducing momentum to the basic equation of CCSD given in \cite{GS95}, we have simultaneous equations similarly to \cite{HPTB04} for single-excitation amplitudes as 
 \begin{align}
     0 = &  f_{p \bar{\bm{k}}}^{g \bar{\bm{k}}}
             + \sum_c \tilde{\mathcal{F}}^{g \bar{\bm{k}} c \bar{\bm{k}}} t_{p \bar{\bm{k}}}^{c \bar{\bm{k}}}
             - \sum_r t_{r \bar{\bm{k}}}^{g \bar{\bm{k}}}  \tilde{\mathcal{F}}_{r \bar{\bm{k}} p \bar{\bm{k}}}
             + \sum_{n, f, \hat{\bm{k}}} \tilde{\mathcal{F}}_{n \hat{\bm{k}}}^{f \hat{\bm{k}}} t_{p \bar{\bm{k}} n \hat{\bm{k}}}^{g \bar{\bm{k}} f \hat{\bm{k}}}  \notag \\
           &+ \sum_{r, c, \hat{\bm{k}}} t_{r \hat{\bm{k}}}^{c \hat{\bm{k}}} \bra{g \bar{\bm{k}} r \hat{\bm{k}}} \ket{p \bar{\bm{k}} c \hat{\bm{k}}}_{voov}  \notag \\
           &- \frac{1}{2} \sum_{m, \bm{k}_m, n, \bm{k}_n, f}  t_{m \bm{k}_m   n \bm{k}_n}^{g \bar{\bm{k}} f \bm{k}_f} \bra{m \bm{k}_m n \bm{k}_n} \ket{p \bar{\bm{k}} f \bm{k}_f}_{ooov}  \notag \\
           &+ \frac{1}{2} \sum_{n, \bm{k}_n, e, \bm{k}_e, f}  t_{p \bar{\bm{k}} n \bm{k}_n}^{e \bm{k}_e f \bm{k}_f} \bra{g \bar{\bm{k}} n \bm{k}_n} \ket{e \bm{k}_e f \bm{k}_f}_{vovv},  \label{ori_EqSingle}
 \end{align}
 and for double-excitation amplitudes as
 \begin{align}
     0 = &  \bra{a \bm{k}_a b \bm{k}_b} \ket{i \bm{k}_i j \bm{k}_j}_{vvoo}  \notag  \\ 
           &+ P_- (ab) \sum_f  t_{i \bm{k}_i j \bm{k}_j}^{a \bm{k}_a f \bm{k}_b}   \left( \tilde{\mathcal{F}}^{b \bm{k}_b f \bm{k}_b} - \frac{1}{2} \sum_r  t_{r \bm{k}_b}^{b \bm{k}_b} \tilde{\mathcal{F}}_{r \bm{k}_b}^{f \bm{k}_b} \right)  \notag \\
           &-  P_- (ij)   \sum_n t_{i \bm{k}_i n \bm{k}_j}^{a \bm{k}_a b \bm{k}_b} \left( \tilde{\mathcal{F}}_{n \bm{k}_j j \bm{k}_j}   + \frac{1}{2} \sum_c t_{j \bm{k}_j}^{c \bm{k}_j}    \tilde{\mathcal{F}}_{n \bm{k}_j}^{c \bm{k}_j}  \right)  \notag \\
           &+ \frac{1}{2} \sum_{m, \bm{k}_m, n} \tau_{m \bm{k}_m n \bm{k}_n}^{a \bm{k}_a b \bm{k}_b} \tilde{\mathcal{W}}_{m \bm{k}_m n \bm{k}_n i \bm{k}_i j \bm{k}_j}
             + \frac{1}{2} \sum_{e, \bm{k}_e, f}    \tilde{\mathcal{W}}^{a \bm{k}_a b \bm{k}_b e \bm{k}_e f \bm{k}_f} \tau_{i \bm{k}_i j \bm{k}_j}^{e \bm{k}_e f \bm{k}_f}   \notag \\
           &+ P_- (ab) P_- (ij) \left(   \sum_{n, \bm{k}_n, f} t_{i \bm{k}_i n \bm{k}_n}^{a \bm{k}_a f \bm{k}_f} \tilde{\mathcal{W}}_{j \bm{k}_j n \bm{k}_n}^{b \bm{k}_b f \bm{k}_f} \right.   \notag \\
           &~~~~~~~~~~~~~~~~~~~~~~~~~~~~~~~   \left. - \sum_{r, c} t_{i \bm{k}_i}^{c \bm{k}_i} t_{r \bm{k}_a}^{a \bm{k}_a} \bra{r \bm{k}_a b \bm{k}_b} \ket{c \bm{k}_i j \bm{k}_j}_{ovvo} \right)  \notag \\
           &+ P_- (ij)    \sum_c \bra{a \bm{k}_a b \bm{k}_b} \ket{c \bm{k}_i j \bm{k}_j}_{vvvo} t_{i \bm{k}_i}^{c \bm{k}_i}  \notag \\
           &-  P_- (ab) \sum_r  t_{r \bm{k}_a}^{a \bm{k}_a} \bra{r \bm{k}_a b \bm{k}_b} \ket{c \bm{k}_i j \bm{k}_j}_{ovoo},  \label{ori_EqDouble}
 \end{align}
 where $P_- ()$ is an operator, $\tilde{\mathcal{W}}$ and $\tilde{\mathcal{F}}$ are intermediates and $\tau$ are effective double-excitation amplitudes.
 The operator $P_- ()$ acts on a quantity $Z$ as
 \begin{equation}
     P_- (rs) Z(\cdots r \bm{k}_r s \bm{k}_s \cdots ) = Z(\cdots r \bm{k}_r s \bm{k}_s \cdots ) - Z(\cdots s \bm{k}_s r \bm{k}_r \cdots ).
 \end{equation}
 The intermediates $\tilde{\mathcal{W}}$ are defined as
 \begin{align}
      \tilde{\mathcal{W}}_{m \bm{k}_m n \bm{k}_n i \bm{k}_i j \bm{k}_j}  =&                                                                                                   \bra{m\bm{k}_m n \bm{k}_n} \ket{i \bm{k}_i j \bm{k}_j}_{oooo}  \notag \\
                                                                                                                 &+ P_- (ij) \sum_c t_{j \bm{k}_j}^{c \bm{k}_j}                                \bra{m\bm{k}_m n \bm{k}_n} \ket{i \bm{k}_i c \bm{k}_j}_{ooov}  \notag \\
                                                                                                                 &+ \frac{1}{4} \sum_{e, \bm{k}_e, f} \tau_{i \bm{k}_i j \bm{k}_j}^{e \bm{k}_e f \bm{k}_f} \bra{m\bm{k}_m n \bm{k}_n} \ket{e \bm{k}_e f \bm{k}_f}_{oovv},   \\
      \tilde{\mathcal{W}}^{a \bm{k}_a b \bm{k}_b e \bm{k}_e f \bm{k}_f}  =&                                                                                                   \bra{a \bm{k}_a b \bm{k}_b} \ket{e \bm{k}_e f \bm{k}_f}_{vvvv}  \notag \\
                                                                                                                 &- P_- (ab) \sum_r t_{r \bm{k}_b}^{b \bm{k}_b}                            \bra{a \bm{k}_a r \bm{k}_b} \ket{e \bm{k}_e f \bm{k}_f}_{vovv}  \notag \\
                                                                                                                 &+ \frac{1}{4} \sum_{m, \bm{k}_m, n} \tau_{m\bm{k}_m n \bm{k}_n}^{a \bm{k}_a b \bm{k}_b} \bra{m\bm{k}_m n \bm{k}_n} \ket{e \bm{k}_e f \bm{k}_f}_{oovv},  \\
      \tilde{\mathcal{W}}_{j \bm{k}_j n \bm{k}_n}^{b \bm{k}_b f \bm{k}_f} =&                                                                                                   \bra{n \bm{k}_n b \bm{k}_b} \ket{f \bm{k}_f j \bm{k}_j}_{ovvo}  \notag \\
                                                                                                                 &+ \sum_c t_{j \bm{k}_j}^{c \bm{k}_j}                                            \bra{n \bm{k}_n b \bm{k}_b} \ket{f \bm{k}_f c \bm{k}_j}_{ovvv}  \notag \\
                                                                                                                 &-  \sum_r  t_{r \bm{k}_b}^{b \bm{k}_b}                                         \bra{n \bm{k}_n r \bm{k}_r} \ket{f \bm{k}_f j \bm{k}_j}_{oovo}  \notag \\
                                                                                                                 &-  \frac{1}{2} \sum_{x, \bm{k}_x, y} t_{j \bm{k}_j x \bm{k}_x}^{y \bm{k}_y b \bm{k}_b} \bra{n \bm{k}_n x \bm{k}_x} \ket{f \bm{k}_f y \bm{k}_y}_{oovv}  \notag \\
                                                                                                                 &-  \sum_c  \sum_r t_{j \bm{k}_j}^{c \bm{k}_j} t_{r \bm{k}_b}^{b \bm{k}_b} \bra{n \bm{k}_n r \bm{k}_b} \ket{f \bm{k}_f c \bm{k}_j}_{oovv}.  \label{Def_Woovv}
 \end{align}
 The intermediates $\tilde{\mathcal{F}}$ are defined as
 \begin{align}
     \tilde{\mathcal{F}}_{p \bar{\bm{k}} q \bar{\bm{k}}}     =&                                                           f_{p \bar{\bm{k}} q \bar{\bm{k}}}
                                                                                           + \frac{1}{2} \sum_c                            f_{p \bar{\bm{k}}}^{c \bar{\bm{k}}} t_{q \bar{\bm{k}}}^{c \bar{\bm{k}}}
                                                                                           +                  \sum_{r, c, \hat{\bm{k}}}  t_{r \hat{\bm{k}}}^{c \hat{\bm{k}}} \bra{p \bar{\bm{k}} r \hat{\bm{k}}} \ket{q \bar{\bm{k}} c \hat{\bm{k}}}_{ooov}  \notag \\
                                                                                         &+ \frac{1}{2} \sum_{n, \bm{k}_n, e, \bm{k}_e, f} \tilde{\tau}_{q \bar{\bm{k}} n \bm{k}_n}^{e \bm{k}_e f \bm{k}_f} \bra{p \bar{\bm{k}} n \bm{k}_n} \ket{e \bm{k}_e f \bm{k}_f}_{oovv} ,  \label{ori_F_oo}  \\
     \tilde{\mathcal{F}}^{g \bar{\bm{k}} h \bar{\bm{k}}}     =&                                                           f_{g \bar{\bm{k}} h \bar{\bm{k}}}
                                                                                           - \frac{1}{2}  \sum_r                             f_{r \bar{\bm{k}}}^{h \bar{\bm{k}}} t_{r \bar{\bm{k}}}^{g \bar{\bm{k}}}
                                                                                           +                  \sum_{r, c, \hat{\bm{k}}}  t_{r \hat{\bm{k}}}^{c \hat{\bm{k}}} \bra{g \bar{\bm{k}} r \hat{\bm{k}}} \ket{h \bar{\bm{k}} c \hat{\bm{k}}}_{ooov}  \notag \\
                                                                                         &- \frac{1}{2} \sum_{m, \bm{k}_m, n, \bm{k}_n, f} \tilde{\tau}_{m \bm{k}_m n \bm{k}_n}^{g \bar{\bm{k}} f \bm{k}_f} \bra{m \bm{k}_m n \bm{k}_n} \ket{h \bar{\bm{k}} f \bm{k}_f}_{oovv} ,  \label{ori_F_vv} \\
     \tilde{\mathcal{F}}_{p \bar{\bm{k}}}^{g \bar{\bm{k}}}  =&                                                           f_{p \bar{\bm{k}}}^{g \bar{\bm{k}}}
                                                                                           +                  \sum_{r, c, \hat{\bm{k}}}  t_{r \hat{\bm{k}}}^{c \hat{\bm{k}}} \bra{p \bar{\bm{k}} r \hat{\bm{k}}} \ket{g \bar{\bm{k}} c \hat{\bm{k}}}_{oovv},  \label{ori_F_ov}
 \end{align}
 where $\tilde{\tau}$ are effective double-excitation amplitudes
 \begin{equation}
     \tilde{\tau}_{i \bm{k}_i j \bm{k}_j}^{a \bm{k}_a b \bm{k}_b} = t_{i \bm{k}_i j \bm{k}_j}^{a \bm{k}_a b \bm{k}_b} + \frac{1}{4} P_- (ab) P_- (ij) t_{i \bm{k}_i}^{a \bm{k}_a} t_{ j \bm{k}_j}^{b \bm{k}_b}.
 \end{equation} 
 The effective double excitation amplitudes are given as
 \begin{equation}
     \tau_{i \bm{k}_i j \bm{k}_j}^{a \bm{k}_a b \bm{k}_b} = t_{i \bm{k}_i j \bm{k}_j}^{a \bm{k}_a b \bm{k}_b} + \frac{1}{2} P_- (ab) P_- (ij) t_{i \bm{k}_i}^{a \bm{k}_a} t_{ j \bm{k}_j}^{b \bm{k}_b}.
 \end{equation}

  The number of equations in the simultaneous equations for single-excitation amplitudes is $N_{\textrm{occ}} N_{\textrm{vir}} N_{\bm{k}}$,
  where $N_{\textrm{occ}}$, $N_{\textrm{vir}}$ and $N_{\bm{k}}$ are the number of occupied spin orbitals, that of virtual spin orbitals, and that of possible momenta of an electron, respectively.
  Double-excitation amplitudes and antisymmetrized two-electron integrals may have a nonzero value only when the four momentums which identify these quantities satisfy conservation law of momentum.
  Then, the number of equations in the simultaneous equations for double-excitation amplitudes is $N_{\textrm{occ}}^2 N_{\textrm{vir}}^2 N_{\bm{k}}^3$.

  \section{Outlines and rearranged simultaneous equations of the presented method}  \label{OutlineBasicEq}

  In this section, we  describe outlines of the presented method and give its basic equations.
  The basic equations are obtained by rearranging the basic method shown in the previous section.
  
  In Section \ref{MethodOutline}, outlines of the presented method is described.
  In Section \ref{MethodPreconditions}, preconditions in the presented method are given.
  In Section \ref{RearrangedEq}, rearranged simultaneous equations for the presented method are given.

  \subsection{Outlines of the presented method}  \label{MethodOutline}

  In numerical computation of the CCSD in periodic systems, handling of quantities which have eight indices may become a problem.
  In a na\"ive parallel computing implementation, we will be suffered from a large amount of communication of such quantities between parallel computing processes.
  To avoid this problem,
  we use key ideas and some techniques for implementation developed for algorithm for the Higher Order Tensor Renormalization Group method (HOTRG) \cite{XCQZYX12}
  which is described in a paper in preparation by T. Y. and T. S. \cite{YS19} to be shown in another place.
  Akiyama, Kuramashi, T. Y. and Yoshimura \cite{AKYY19} argue phase transition of the four-dimensional Ising model using codes basically based on this algorithm.
  The key ideas for the HOTRG are to utilize process numbers in parallel computing to identify some indices of tensors and these indices should {\it not be contracted} during considering step in computation.
  The quantities which appear the basic equation of the CCSD in a periodic system can be regarded as tensor elements.
  Then, in the method we present in this paper, the process numbers are utilized to identify two of the eight indices.
  Slightly different from the case of the HOTRG, there are cases such that there exists only one index which is not contracted during considering step in computation.
  Then, a principal for choice of indices which are identified through a process number in the CCSD in periodic systems
  is that we should {\it preferentially} choose indices which {\it are not contracted} during considering step in computation.
  In addition to this principal, we consider conservation law of momentum since this consideration gives a good perspective for design of a method for communication between parallel computing processes.
  As a conclusion, two indices which represent momentum of an electron are identified though a process number.
  In the presented method, the $k$ points are represented by one of the numbers $0, 1, ..., N_{\bm{k}} - 1$.
  Then, our method prepares $N_{\bm{k}}^2$ processes.
  When the process number of a process is represented as $\alpha + \beta N_{\bm{k}}$ $( \alpha , \beta = 0, 1, ..., N_{\bm{k}} - 1 )$,
  the identified momentums $\bm{k}_1$ and $\bm{k}_2$ are $\bm{k}_1 = \alpha$ and $\bm{k}_2 = \beta$.
  Some techniques which are not used in the HOTRG are introduced in this paper.
  
  We consider computational cost and required memory space.
  For simplicity, we consider a case such that the numbers of occupied and virtual bands are equal.
  Let us denote these numbers by $N_{\textrm{band}}$.
  In a na\"ive implementation, computational cost and required memory space are $O ( N_{\textrm{band}}^6 N_{\bm{k}}^4 )$ and $O ( N_{\textrm{band}}^4 N_{\bm{k}}^3 )$, respectively, because of conservation law of momentum.
  In the presented method, computational cost is $O ( N_{\textrm{band}}^6 N_{\bm{k}}^2 )$ and required memory space in each process is $O ( N_{\textrm{band}}^4 N_{\bm{k}} )$.

  \subsection{Preconditions}  \label{MethodPreconditions}

  In this section, preconditions in the presented method are described.

  \subsubsection{Processes in parallel computing}
  
\begin{figure}[H]
      \begin{center}
         \includegraphics[width=0.3\textwidth]{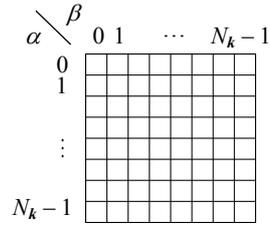}
         \caption{Processes}
         \label{Fig_Processes}
      \end{center}
  \end{figure}

  As mentioned in outlines, $N_k^2$ processes are used in parallel computing.
  A process number can be represented by two integers as $\alpha + \beta N_{\bm{k}}$ $( \alpha , \beta = 0, 1, ..., N_{\bm{k}} - 1 )$.
  Each of $\alpha$ and $\beta$ identifies the momentum of an electron.
  An image shown in Fig. \ref{Fig_Processes} may help understanding of the presented method.
  Each box represents a process.
  Rows and columns correspond to $\alpha$ and $\beta$, respectively.
  A box at the intersection of each row and each column represents a process whose number is $\alpha + \beta N_{\bm{k}}$.

\begin{figure}[H]
      \begin{center}
         \includegraphics[width=0.5\textwidth]{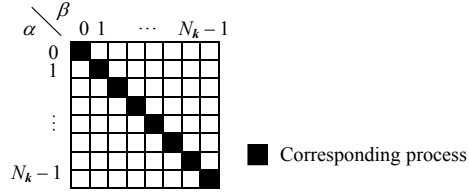}
         \caption{The diagonal processes}
         \label{Fig_Diag_Process}
      \end{center}
  \end{figure}

  \noindent
  During computing, there are cases such that we consider only processes whose process numbers are represented as $p + p N_{\bm{k}}$ $( p = 0, 1, ..., N_{\bm{k}} - 1 )$.
  See Fig. \ref{Fig_Diag_Process}.
  Let us call these processes ``the diagonal processes''.

\begin{figure}[H]
      \begin{center}
         \includegraphics[width=0.6\textwidth]{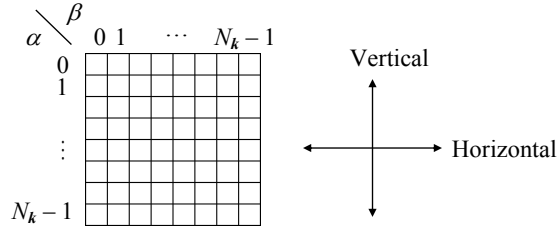}
         \caption{Directions}
         \label{Fig_direction}
      \end{center}
  \end{figure}

  \noindent
  When we represent process numbers as $\alpha + \beta N_{\bm{k}}$,
  there are cases such that we want to handle processes which have the same $\alpha$ or $\beta$ as a group.
  For this purpose, let us introduce ``direction'' into the diagram shown in shown in Fig. \ref{Fig_Processes}.
  Horizontal and vertical directions are introduced as shown in Fig. \ref{Fig_direction}.

  \subsubsection{Inputs and outputs}  \label{IO}
  
  Inputs are the known quantities and the initial values of the unknown quantities of simultaneous equations.
  The matrix elements of the Fock operator $f_{p \bar{\bm{k}}}^{g \bar{\bm{k}}}$, $f_{p \bar{\bm{k}} q \bar{\bm{k}}}$ and $f^{g \bar{\bm{k}} h \bar{\bm{k}}}$ $( \bar{\bm{k}} = 0, 1, ..., N_{\bm{k}} - 1 )$ 
  are stored in the diagonal process whose number is $\bar{\bm{k}} + \bar{\bm{k}} N_{\bm{k}}$.
  See Fig. \ref{Fig_Diag_Process}.
  Antisymmetrized two-electron integrals are distributed to each process according to two of four indices which represent momentum of an electron.
  There are plural choices for such two indices and there is no necessity that one integral is distributed to only one process.
  The integrals are distributed to each process as shown in Table \ref{Table_TEI}.
  In the cases such that the integrals are distributed in plural ways,
  superscripts are attached to the integrals for identification.
  The symbol ``$o$" (``$v$") means that corresponding index represents an occupied (a virtual) spin orbital.
  \begin{table}[h]
    \begin{center}
      \caption{Distribution of antisymmetrized two-electron integrals to each process}  \label{Table_TEI}
      \begin{tabular}{| c | c |}
        \hline
        Integrals  &  Process number   \\
        \hline \hline
        $\bra{\alpha \bm{k}_{\alpha} \beta \bm{k}_{\beta}} \ket{\gamma \bm{k}_{\gamma} \delta \bm{k}_{\delta}}_{oooo}$              &  $\bm{k}_{\gamma} + \bm{k}_{\delta} N_{\bm{k}}$  \\
        \hline
        $\bra{\alpha \bm{k}_{\alpha} \beta \bm{k}_{\beta}} \ket{\gamma \bm{k}_{\gamma} \delta \bm{k}_{\delta}}_{ooov}^{[oo]}$    &  $\bm{k}_{\alpha} + \bm{k}_{\gamma} N_{\bm{k}}$  \\
        \hline
        $\bra{\alpha \bm{k}_{\alpha} \beta \bm{k}_{\beta}} \ket{\gamma \bm{k}_{\gamma} \delta \bm{k}_{\delta}}_{ooov}^{[ov]}$    &  $\bm{k}_{\gamma} + \bm{k}_{\delta} N_{\bm{k}}$  \\
        \hline
        $\bra{\alpha \bm{k}_{\alpha} \beta \bm{k}_{\beta}} \ket{\gamma \bm{k}_{\gamma} \delta \bm{k}_{\delta}}_{oovv}^{[oo]}$    &  $\bm{k}_{\alpha} + \bm{k}_{\beta} N_{\bm{k}}$  \\
        \hline
        $\bra{\alpha \bm{k}_{\alpha} \beta \bm{k}_{\beta}} \ket{\gamma \bm{k}_{\gamma} \delta \bm{k}_{\delta}}_{oovv}^{[ov]}$    &  $\bm{k}_{\alpha} + \bm{k}_{\gamma} N_{\bm{k}}$  \\
        \hline
        $\bra{\alpha \bm{k}_{\alpha} \beta \bm{k}_{\beta}} \ket{\gamma \bm{k}_{\gamma} \delta \bm{k}_{\delta}}_{oovv}^{[vv]}$    &  $\bm{k}_{\gamma} + \bm{k}_{\delta} N_{\bm{k}}$  \\
        \hline
        $\bra{\alpha \bm{k}_{\alpha} \beta \bm{k}_{\beta}} \ket{\gamma \bm{k}_{\gamma} \delta \bm{k}_{\delta}}_{ovvo}$             &  $\bm{k}_{\delta} + \bm{k}_{\beta} N_{\bm{k}}$  \\
        \hline
        $\bra{\alpha \bm{k}_{\alpha} \beta \bm{k}_{\beta}} \ket{\gamma \bm{k}_{\gamma} \delta \bm{k}_{\delta}}_{ovvv}^{[ov]}$    &  $\bm{k}_{\alpha} + \bm{k}_{\beta} N_{\bm{k}}$  \\
        \hline
        $\bra{\alpha \bm{k}_{\alpha} \beta \bm{k}_{\beta}} \ket{\gamma \bm{k}_{\gamma} \delta \bm{k}_{\delta}}_{ovvv}^{[vv]}$    &  $\bm{k}_{\beta} + \bm{k}_{\delta} N_{\bm{k}}$  \\
        \hline
        $\bra{\alpha \bm{k}_{\alpha} \beta \bm{k}_{\beta}} \ket{\gamma \bm{k}_{\gamma} \delta \bm{k}_{\delta}}_{vvvv}          $    &  $\bm{k}_{\alpha} + \bm{k}_{\beta} N_{\bm{k}}$  \\
        \hline
      \end{tabular}
    \end{center}
  \end{table}
  
  \noindent
  We can set the initial values of the unknown quantities of simultaneous equations, single-excitation and double-excitation amplitudes, to arbitrary values.
  For setting the initial values of single-excitation and double-excitation amplitudes in our implementation, we rewrite the equations to be solved to a form $t=f(t)$, where $t$ represents the amplitudes collectively.
  We set the initial amplitudes $t_0$ to the right-hand side with $t = 0$, that is,  $t_0$ = $f(0)$.
  Outputs are these unknown quantities.

  \subsubsection{Parallel computing of the right-hand sides of simultaneous equations}  \label{RHS_abstract}
  
\begin{figure}[H]
      \begin{center}
         \includegraphics[width=0.8\textwidth]{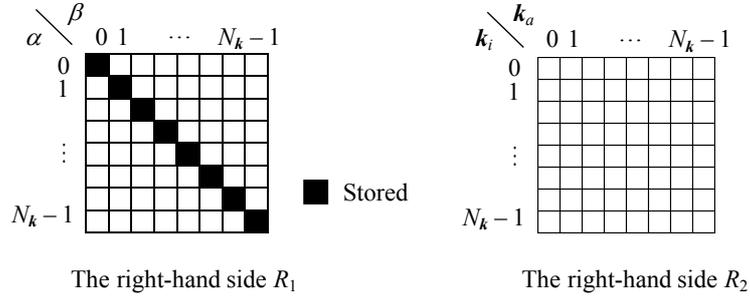}
         \caption{Parallel computing of the right-hand sides of the simultaneous equations}
         \label{Fig_RHS}
      \end{center}
  \end{figure}

  In the presented method, the basic equations are rearranged to another form of simultaneous equations described below.
  The rearranged simultaneous equations for single-excitation amplitudes (\ref{EqSingle}) shown below can be conceptionally written as
  \begin{equation}  \label{Eq_R1}
      0 = ( R_1 )_{g \bar{\bm{k}}}^{p \bar{\bm{k}}}.
  \end{equation}
  Those for double-excitation amplitudes (\ref{EqDouble}) shown below can be conceptionally written as
  \begin{equation}  \label{Eq_R2}
      0 = ( R_2 )_{i {\bm{k}}_i j {\bm{k}}_j}^{a {\bm{k}}_a b {\bm{k}}_b}.
  \end{equation}
  Intermediate results during computation and the final results in computation of $R_1$ are stored in the diagonal process whose process number is $\bar{\bm{k}} + \bar{\bm{k}} N_{\bm{k}}$.
  Those in computation of $R_2$ are stored in the process whose process number is $\bm{k}_i + \bm{k}_a N_{\bm{k}}$.
  See Fig. \ref{Fig_RHS}.

  \subsubsection{Distribution of single-excitation and double-excitation amplitudes to processes}  \label{DistrbutionAmp}

\begin{figure}[H]
      \begin{center}
         \includegraphics[width=0.3\textwidth]{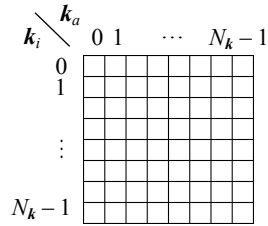}
         \caption{Processes in which double-excitation amplitudes are stored}
         \label{Fig_t2_stored}
      \end{center}
  \end{figure}

  Single-excitation amplitudes $t_{p \bar{\bm{k}}}^{g \bar{\bm{k}}}$ are stored in all the processes.
  Those of double-excitation amplitudes $t_{i {\bm{k}}_i j {\bm{k}}_j}^{a {\bm{k}}_a b {\bm{k}}_b}$ are distributedly stored in a process according to the momentums ${\bm{k}}_i$ and ${\bm{k}}_a$.
  They are stored in a process whose process number is $\bm{k}_i + \bm{k}_a N_{\bm{k}}$.
  See Fig. \ref{Fig_t2_stored}.

  \subsubsection{Conservation law of momentum}

  Throughout the present algorithm, the momentum conservation in a periodic system has to be taken into account.
  Specifically, the equivalence of $\bm{k}_1 + \bm{k}_2$ and $\bm{k}_3 + \bm{k}_4$ means that there exists a reciprocal lattice vector $\bm{G}$ such that $\bm{k}_1 + \bm{k}_2 = \bm{k}_3 + \bm{k}_4 + \bm{G}$.
  If one adopt a regular mesh containing $\bm{k}= 0$ in the reciprocal space as usual, there will be no difficulty in finding the momentum from given three momenta.

 \subsection{Rearrangement of the basic equations in the CCSD in a periodic system}  \label{RearrangedEq}
 
  In this section, we present simultaneous equations which are suitable for parallel computing in the presented method.
  These simultaneous equations are obtained by rearranging the basic equations and the intermediates shown in Section \ref{BasicEqs}.
  The order of items in the right-hand sides in the simultaneous equations in Section \ref{BasicEqs} for single-excitation and double-excitation amplitudes are changed.
  A few intermediates are newly introduced into the rearranged equations.
  Some of intermediates used in Section \ref{BasicEqs} are decomposed, namely, are not used.
  
  For antisymmetrized two-electron integrals, we apply the following relations
  \begin{equation}
      \bra{p \bm{k}_p q \bm{k}_q} \ket{r \bm{k}_r s \bm{k}_s} = \bra{r \bm{k}_r s \bm{k}_s} \ket{p \bm{k}_p q \bm{k}_q}^*
  \end{equation}
  and
  \begin{align}
           \bra{p \bm{k}_p q \bm{k}_q} \ket{r \bm{k}_r s \bm{k}_s}
      &= - \bra{q \bm{k}_q p \bm{k}_p} \ket{r \bm{k}_r s \bm{k}_s}  \notag \\
      &= - \bra{p \bm{k}_p q \bm{k}_q} \ket{s \bm{k}_s r \bm{k}_r}
        =   \bra{q \bm{k}_q p \bm{k}_p} \ket{s \bm{k}_s r \bm{k}_r}.
  \end{align}

  The rearrangement of the basic equations are described as follows.
  In Section \ref{introduce_rho}, intermediates $\check{\rho}$ and $\hat{\rho}$ are newly introduced.
  In section \ref{arrange_F}, rearrangement of the intermediates $\tilde{\mathcal{F}}$ is described.
  In Section \ref{Sec_EqSingle} and \ref{Sec_EqSingle}, rearranged simultaneous equations for single-excitation and double-excitation amplitudes are given, respectively.
  
  \subsubsection{Introduction of new intermediates $\check{\rho}$ and $\hat{\rho}$}  \label{introduce_rho}

  Intermediates $\check{\rho}$ and $\hat{\rho}$ are newly introduced.
  In our parallel computing method, they depend on a process in which they are computed.
  When a process number is expressed as $\bm{k}_{\alpha} + N_{\bm{k}} \bm{k}_{\beta}$ $( \bm{k}_{\alpha}, \bm{k}_{\beta} = 0, 1, ..., N_{\bm{k}} - 1 )$,
  definitions of these intermediates are
  \begin{gather}
      \check{\rho}_p^g ( \bm{k}_{\alpha} ) = t_{p \bm{k}_{\alpha}}^{g \bm{k}_{\alpha}},  \label{def_hat_rho}  \\
      \hat{\rho}_p^g ( \bm{k}_{\beta} ) = t_{p\bm{k}_{\beta}}^{g \bm{k}_{\beta}}.  \label{def_check_rho}
  \end{gather}

  \subsubsection{Rearrangement of the intermediates $\tilde{\mathcal{F}}$}  \label{arrange_F}

  The intermediates $\tilde{\mathcal{F}}$ are rearranged from (\ref{ori_F_oo}), (\ref{ori_F_vv}) and (\ref{ori_F_ov}) as
  \begin{align}
       \tilde{\mathcal{F}}_{p \bar{\bm{k}} q \bar{\bm{k}}}
       =&  f_{p \bar{\bm{k}} q \bar{\bm{k}}}
           + \frac{1}{2} \sum_c \check{\rho}_q^c ( \bar{\bm{k}} ) \left( f_{p \bar{\bm{k}}}^{c \bar{\bm{k}}} - \tilde{\mathcal{K}}_{p \bar{\bm{k}}}^{c \bar{\bm{k}}} \right)  \notag  \\
         &+ \sum_{r, c, \hat{\bm{k}}} t_{r \hat{\bm{k}}}^{c \hat{\bm{k}}}  \bra{p \bar{\bm{k}} r \hat{\bm{k}}} \ket{q \bar{\bm{k}} c \hat{\bm{k}}}_{ooov}^{[oo]}
           + \sum_{\bm{k}_e} \tilde{\mathcal{Z}}_{p \bar{\bm{k}} q \bar{\bm{k}}}^{\bm{k}_e},  \label{def_F_oo}  \\
       \tilde{\mathcal{F}}^{g \bar{\bm{k}} h \bar{\bm{k}}}
       =&  f^{g \bar{\bm{k}} h \bar{\bm{k}}}
           - \frac{1}{2} \sum_r  \hat{\rho}_r^g ( \bar{\bm{k}} ) \left( f_{r \bar{\bm{k}}}^{h \bar{\bm{k}}} + \tilde{\mathcal{K}}_{r \bar{\bm{k}}}^{h \bar{\bm{k}}} \right)   \notag  \\
         &+ \sum_{r, c, \hat{\bm{k}}} t_{r \hat{\bm{k}}}^{c \hat{\bm{k}}} \bra{r \hat{\bm{k}} g \bar{\bm{k}}} \ket{c \hat{\bm{k}} h \bar{\bm{k}}}_{ovvv}^{[vv]}
            -  \sum_{\bm{k}_m} \tilde{\mathcal{Z}}_{\bm{k}_m}^{g \bar{\bm{k}} h \bar{\bm{k}}},  \label{def_F_vv}  \\
       \tilde{\mathcal{F}}_{p \bar{\bm{k}}}^{g \bar{\bm{k}}} =& f_{p \bar{\bm{k}}}^{g \bar{\bm{k}}}  + \tilde{\mathcal{K}}_{p \bar{\bm{k}}}^{g \bar{\bm{k}}},  \label{def_F_ov}
  \end{align}
  where $ \tilde{\mathcal{Z}}$ and $ \tilde{\mathcal{K}}$ are newly introduced intermediates defined as
  \begin{gather}
      \tilde{\mathcal{Z}}_{p \bar{\bm{k}} q \bar{\bm{k}}}^{\bm{k}_e} = \frac{1}{2} \sum_e \sum_{n, \bm{k}_n, f} t_{q \bar{\bm{k}} n \bm{k}_n}^{e \bm{k}_e f \bm{k}_f} \bra{p \bar{\bm{k}} n \bm{k}_n} \ket{e \bm{k}_e f \bm{k}_f}_{oovv}^{[ov]}  ,  \label{def_Z_oo}  \\
      \tilde{\mathcal{Z}}_{\bm{k}_m}^{g \bar{\bm{k}} h \bar{\bm{k}}} = \frac{1}{2} \sum_m \sum_{n, \bm{k}_n, f} t_{m \bm{k}_m n \bm{k}_n}^{g \bar{\bm{k}} f \bm{k}_f} \bra{m \bm{k}_m n \bm{k}_n} \ket{h \bar{\bm{k}} f \bm{k}_f}_{oovv}^{[ov]}  ,  \label{def_Z_vv}  \\
      \tilde{\mathcal{K}}_{x \bar{\bm{k}}}^{y \bar{\bm{k}}} = \sum_{r, c, \hat{\bm{k}}} t_{r \hat{\bm{k}}}^{c \hat{\bm{k}}} \bra{x \bar{\bm{k}} r  \hat{\bm{k}}} \ket{y \bar{\bm{k}} c  \hat{\bm{k}}}_{oovv}^{[ov]}.  \label{def_K}
  \end{gather}

  \subsubsection{The rearranged simultaneous equations for single-excitation amplitudes}  \label{Sec_EqSingle}
  
  The rearranged simultaneous equations for single-excitation amplitudes are obtained from (\ref{ori_EqSingle}) as
  \begin{align}  \label{EqSingle}
      0 =&   f_{p \bar{\bm{k}}}^{g \bar{\bm{k}}}
              + \sum_c \check{\rho}_p^c ( \bar{\bm{k}} )    \tilde{\mathcal{F}}^{g \bar{\bm{k}} c \bar{\bm{k}}}
              -  \sum_r \hat{\rho}_r^g ( \bar{\bm{k}} )     \tilde{\mathcal{F}}_{r \bar{\bm{k}} p \bar{\bm{k}}}
              + \sum_{n, f, \hat{\bm{k}}} t_{p \bar{\bm{k}} n \hat{\bm{k}}}^{g \bar{\bm{k}} f \hat{\bm{k}}} \tilde{\mathcal{F}}_{n \hat{\bm{k}}}^{f \hat{\bm{k}}}  \notag \\
           &+ \sum_{r, c, \hat{\bm{k}}} t_{r \hat{\bm{k}}}^{c \hat{\bm{k}}} \bra{r \hat{\bm{k}} g \bar{\bm{k}}} \ket{c \hat{\bm{k}} p \bar{\bm{k}}}_{ovvo}
              - \sum_{\bm{k}_m} \tilde{\mathcal{L}}_{p \bar{\bm{k}} \bm{k}_m}^{g \bar{\bm{k}}}
              + \sum_{\bm{k}_e} \tilde{\mathcal{L}}_{p \bar{\bm{k}}}^{g \bar{\bm{k}} \bm{k}_e}.
  \end{align}
  The intermediates $\tilde{\mathcal{L}}$ are newly introduced and defined as
  \begin{gather}
           \tilde{\mathcal{L}}_{p \bar{\bm{k}} \bm{k}_m}^{g \bar{\bm{k}}} = \frac{1}{2}  \sum_m \sum_{n, \bm{k}_n, f}  t_{m \bm{k}_m   n \bm{k}_n}^{g \bar{\bm{k}} f \bm{k}_f} \bra{m \bm{k}_m n \bm{k}_n} \ket{p \bar{\bm{k}} f \bm{k}_f}_{ooov}^{[oo]},  \label{def_L_oo}  \\
           \tilde{\mathcal{L}}_{p \bar{\bm{k}}}^{g \bar{\bm{k}} \bm{k}_e}  = \frac{1}{2}  \sum_e  \sum_{n, \bm{k}_n, f}  t_{p \bar{\bm{k}} n \bm{k}_n}^{e \bm{k}_e f \bm{k}_f} \bra{n \bm{k}_n g \bar{\bm{k}}} \ket{f \bm{k}_f e \bm{k}_e}_{ovvv}^{[vv]}.  \label{def_L_vv}
  \end{gather}

  \subsubsection{The rearranged simultaneous equations for double-excitation amplitudes}  \label{Sec_EqDouble}

  We rearrange the right-hand side of the rearranged simultaneous equations for double-excitation amplitudes (\ref{ori_EqDouble}).
  Numerical computation of the rearranged right-hand side is divided into four parts.
  Among each parts, ways of identification of momentums through a process number is different.
  Then, the rearranged simultaneous equations is expressed as
  \begin{equation}  \label{EqDouble}
     0 =    \bar{\mathcal{A}}_{i \bm{k}_i j \bm{k}_j}^{a \bm{k}_a b \bm{k}_b}
           + \check{\mathcal{A}}_{i \bm{k}_i j \bm{k}_j}^{a \bm{k}_a b \bm{k}_b}
           + \tilde{\mathcal{A}}_{i \bm{k}_i j \bm{k}_j}^{a \bm{k}_a b \bm{k}_b}
           + \hat{\mathcal{A}}_{i \bm{k}_i j \bm{k}_j}^{a \bm{k}_a b \bm{k}_b}.
  \end{equation}

  For the first term $\bar{\mathcal{A}}_{i \bm{k}_i j \bm{k}_j}^{a \bm{k}_a b \bm{k}_b}$, the momentums $\bm{k}_i$ and $\bm{k}_a$ are identified through a process number.
  This term is given as
  \begin{align}  \label{A_1st}
            \bar{\mathcal{A}}_{i \bm{k}_i j \bm{k}_j}^{a \bm{k}_a b \bm{k}_b}
      =& P_-(ij) P_-(ab) \tilde{\mathcal{C}}_{i \bm{k}_i j \bm{k}_j}^{a \bm{k}_a b \bm{k}_b}
          + \left( \bra{i \bm{k}_i j \bm{k}_j} \ket{a \bm{k}_a b \bm{k}_b}_{oovv} \right) ^* \notag \\
        &+ \sum_{s, d} \left( t_{s \bm{k}_b}^{b \bm{k}_b} t_{j \bm{k}_j}^{d \bm{k}_j} \sum_{r, c} \tilde{\zeta}_{ri}^{ac} ( \bm{k}_i, \bm{k}_a ) \bra{r \bm{k}_a s \bm{k}_b} \ket{c \bm{k}_i d \bm{k}_j}_{oovv}^{[ov]} \right) .
  \end{align}
  Intermediates $\tilde{\mathcal{C}}_{i \bm{k}_i j \bm{k}_j}^{a \bm{k}_a b \bm{k}_b}$ and $\tilde{\zeta}$ are newly introduced.
  Similarly to the intermediates $\check{\rho}$ and $\hat{\rho}$, the intermediate $\tilde{\zeta}$ depends on a process.
  When a process number is expressed as $\bm{k}_{\alpha} + N_{\bm{k}} \bm{k}_{\beta}$ $( \bm{k}_{\alpha}, \bm{k}_{\beta} = 0, 1, ..., N_{\bm{k}} - 1 )$,
  the definition of $\tilde{\zeta}$ is
  \begin{equation}  \label{def_zeta}
      \tilde{\zeta}_{p q}^{g h} (  \bm{k}_{\alpha}, \bm{k}_{\beta} ) =  t_{p \bm{k}_{\alpha}}^{g \bm{k}_{\alpha}} t_{q \bm{k}_{\beta}}^{h \bm{k}_{\beta}}.
  \end{equation}
  The intermediate $\tilde{\mathcal{C}}_{i \bm{k}_i j \bm{k}_j}^{a \bm{k}_a b \bm{k}_b}$ is given as
  \begin{equation}  \label{def_C}
      \tilde{\mathcal{C}}_{i \bm{k}_i j \bm{k}_j}^{a \bm{k}_a b \bm{k}_b}
      =   \sum_{n, \bm{k}_n, f} t_{i \bm{k}_i n \bm{k}_n}^{a \bm{k}_a f \bm{k}_f} \tilde{\mathcal{W}}_{j \bm{k}_j n \bm{k}_n}^{b \bm{k}_b f \bm{k}_f}
         - \sum_{r, c} t_{r \bm{k}_b}^{b \bm{k}_b} t_{j \bm{k}_j}^{c \bm{k}_j} \bra{r \bm{k}_b a \bm{k}_a} \ket{c \bm{k}_j i \bm{k}_i}_{ovvo}.
  \end{equation}
  The intermediate $\tilde{\mathcal{W}}_{j \bm{k}_j n \bm{k}_n}^{b \bm{k}_b f \bm{k}_f}$ is rearranged from the original definition in (\ref{Def_Woovv}) and given as
  \begin{align}  \label{rearranged_W}
      \tilde{\mathcal{W}}_{j \bm{k}_j n \bm{k}_n}^{b \bm{k}_b f \bm{k}_f}
       =&- \frac{1}{2}  \sum_{x, y, \bm{k}_y} t_{j \bm{k}_j y \bm{k}_y}^{x \bm{k}_x b \bm{k}_b} \bra{n \bm{k}_n y \bm{k}_y} \ket{f \bm{k}_f x \bm{k}_x}_{oovv}^{[ov]}  \notag \\
         &+ \bra{n \bm{k}_n b \bm{k}_b} \ket{f \bm{k}_f j \bm{k}_j}_{ovvo}
           + \tilde{\mathcal{J}}_{j \bm{k}_j n \bm{k}_n}^{b \bm{k}_b f \bm{k}_f},
  \end{align}
  where $\tilde{\mathcal{J}}_{j \bm{k}_j n \bm{k}_n}^{b \bm{k}_b f \bm{k}_f}$ is a newly introduced intermediate given as
  \begin{align} \label{def_J}
      \tilde{\mathcal{J}}_{j \bm{k}_j n \bm{k}_n}^{b \bm{k}_b f \bm{k}_f}
       =&  \sum_c \hat{\rho}_j^c ( \bm{k}_j )   \bra{n \bm{k}_n b \bm{k}_b} \ket{f \bm{k}_f c \bm{k}_j}_{ovvv}^{[vv]}  \notag  \\
         &- \sum_r \check{\rho}_r^b ( \bm{k}_b ) \bra{r \bm{k}_b n \bm{k}_n} \ket{j \bm{k}_j f \bm{k}_f}_{ooov}^{[oo]}  \notag  \\
         &- \sum_{r, c} \tilde{\zeta}_{rj}^{bc} ( \bm{k}_b, \bm{k}_j ) \bra{r \bm{k}_b n \bm{k}_n} \ket{c \bm{k}_j f \bm{k}_f}_{oovv}^{[ov]} .
  \end{align}

  For the second term $\check{\mathcal{A}}_{i \bm{k}_i j \bm{k}_j}^{a \bm{k}_a b \bm{k}_b}$, the momentums $\bm{k}_i$ and $\bm{k}_j$ are identified through a process number.
  This term is given as
  \begin{align}  \label{A_2nd}
      \check{\mathcal{A}}_{i \bm{k}_i j \bm{k}_j}^{a \bm{k}_a b \bm{k}_b}
     =&   \sum_{r, s} \tilde{\eta}_{r s}^{a b \bm{k}_a \bm{k}_b} \bra{r \bm{k}_a s \bm{k}_b} \ket{i \bm{k}_i j \bm{k}_j}_{oooo}
         + P_-(ij) \check{\mathcal{G}}_{i \bm{k}_i j \bm{k}_j}^{a \bm{k}_a b \bm{k}_b}  \notag \\
       &+ P_-(ab) \sum_f \left( t_{i \bm{k}_i j \bm{k}_j}^{a \bm{k}_a f \bm{k}_b} \left( \tilde{\mathcal{F}}^{b \bm{k}_b f \bm{k}_b} - \frac{1}{2}\sum_r  t_{r \bm{k}_b}^{b \bm{k}_b}  \tilde{\mathcal{F}}_{r \bm{k}_b}^{f \bm{k}_b} \right) \right)  \notag \\
       &+ \sum_{e, \bm{k}_e, f} t_{i \bm{k}_i j \bm{k}_j}^{e \bm{k}_e f \bm{k}_f} \tilde{\mathcal{Y}}^{a \bm{k}_a b \bm{k}_b e \bm{k}_e f \bm{k}_f}.
  \end{align}
  Intermediates $\tilde{\eta}_{r s}^{a b \bm{k}_a \bm{k}_b}$, $\check{\mathcal{G}}_{i \bm{k}_i j \bm{k}_j}^{a \bm{k}_a b \bm{k}_b}$ and $\tilde{\mathcal{Y}}^{a \bm{k}_a b \bm{k}_b e \bm{k}_e f \bm{k}_f}$ are newly introduced.
  The definition of the intermediate $\tilde{\eta}_{r s}^{a b \bm{k}_a \bm{k}_b}$ is
  \begin{equation}
      \tilde{\eta}_{r s}^{a b \bm{k}_a \bm{k}_b} = t_{r \bm{k}_a}^{a \bm{k}_a} t_{s \bm{k}_b}^{b \bm{k}_b}.
  \end{equation}
  That of the intermediate $\check{\mathcal{G}}_{i \bm{k}_i j \bm{k}_j}^{a \bm{k}_a b \bm{k}_b}$ is
  \begin{align}
      \check{\mathcal{G}}_{i \bm{k}_i j \bm{k}_j}^{a \bm{k}_a b \bm{k}_b}
     =&\sum_c \Biggl( \hat{\rho}_j^c ( \bm{k}_j )            \biggl(  \sum_{r, s}  \tilde{\eta}_{r s}^{a b \bm{k}_a \bm{k}_b} \bra{r \bm{k}_a s \bm{k}_b} \ket{i \bm{k}_i c \bm{k}_j}_{ooov}^{[ov]} \biggr. \Biggr.  \notag \\
       &~~~~~~~~~~~~~~~~~~~~~~~~              \Biggl. \biggl. + \left( \bra{i \bm{k}_i c \bm{k}_j} \ket{a \bm{k}_a b \bm{k}_b}_{ovvv}^{[ov]} \right) ^* \biggr) \Biggr) .
  \end{align}
  That of the intermediate $\tilde{\mathcal{Y}}^{a \bm{k}_a b \bm{k}_b e \bm{k}_e f \bm{k}_f}$ is
  \begin{align}  \label{def_Y_vv}
      \tilde{\mathcal{Y}}^{a \bm{k}_a b \bm{k}_b e \bm{k}_e f \bm{k}_f}
      = \frac{1}{2} &          \biggl( P_-(ab) \tilde{\mathcal{V}}^{a \bm{k}_a b \bm{k}_b e \bm{k}_e f \bm{k}_f} - \bra{b \bm{k}_b a \bm{k}_a} \ket{e \bm{k}_e f \bm{k}_f}_{vvvv} \biggr. \notag \\
                          & ~~~~ \biggl. -  \sum_{r, s} \tilde{\zeta}_{sr}^{ba} ( \bm{k}_b, \bm{k}_a ) \bra{s \bm{k}_b r \bm{k}_a} \ket{e \bm{k}_e f \bm{k}_f}_{oovv}^{[oo]}  \biggr) ,
  \end{align}
  where, $\tilde{\mathcal{V}}^{a \bm{k}_a b \bm{k}_b e \bm{k}_e f \bm{k}_f}$ is newly introduced intermediate defined as
  \begin{equation}
      \tilde{\mathcal{V}}^{a \bm{k}_a b \bm{k}_b e \bm{k}_e f \bm{k}_f} = \sum_r \check{\rho}_r^b ( \bm{k}_b ) \bra{r \bm{k}_b a \bm{k}_a} \ket{e \bm{k}_e f \bm{k}_f}_{ovvv}^{[ov]}.
  \end{equation}

  The third term $\tilde{\mathcal{A}}_{i \bm{k}_i j \bm{k}_j}^{a \bm{k}_a b \bm{k}_b}$ is given as
  \begin{equation}  \label{A_3rd}
            \tilde{\mathcal{A}}_{i \bm{k}_i j \bm{k}_j}^{a \bm{k}_a b \bm{k}_b}
            =  \frac{1}{4} \sum_{m, \bm{k}_m, n} t_{m \bm{k}_m n \bm{k}_n}^{a \bm{k}_a b \bm{k}_b} \tilde{\mathcal{X}}_{m \bm{k}_m n \bm{k}_n i \bm{k}_i j \bm{k}_j}
  \end{equation}
  where newly introduced intermediate $\tilde{\mathcal{X}}_{m \bm{k}_m n \bm{k}_n i \bm{k}_i j \bm{k}_j}$ is
  \begin{equation}  \label{def_X}
            \tilde{\mathcal{X}}_{m \bm{k}_m n \bm{k}_n i \bm{k}_i j \bm{k}_j}
            = \sum_{e, \bm{k}_e, f} t_{i \bm{k}_i j \bm{k}_j}^{e \bm{k}_e f \bm{k}_f} \bra{m \bm{k}_m n \bm{k}_n} \ket{e \bm{k}_e f \bm{k}_f}_{oovv}^{[oo]}.
  \end{equation}
  In computation of the intermediate $\tilde{\mathcal{X}}_{m \bm{k}_m n \bm{k}_n i \bm{k}_i j \bm{k}_j}$ in (\ref{def_X}), the momentums $\bm{k}_i$ and $\bm{k}_j$ are continuously identified through a process number.
  In computation of $\tilde{\mathcal{A}}_{i \bm{k}_i j \bm{k}_j}^{a \bm{k}_a b \bm{k}_b}$ in (\ref{A_3rd}), the momentums $\bm{k}_a$ and $\bm{k}_b$ are identified through a process number.

  For the fourth term $\hat{\mathcal{A}}_{i \bm{k}_i j \bm{k}_j}^{a \bm{k}_a b \bm{k}_b}$, the momentums $\bm{k}_a$ and $\bm{k}_b$ are continuously identified through a process number.
  This term is given as
  \begin{align}  \label{A_4th}
      \hat{\mathcal{A}}_{i \bm{k}_i j \bm{k}_j}^{a \bm{k}_a b \bm{k}_b}
      =&  \sum_{m, \bm{k}_m, n} t_{m \bm{k}_m n \bm{k}_n}^{a \bm{k}_a b \bm{k}_b} \tilde{\mathcal{Y}}_{m \bm{k}_m n \bm{k}_n i \bm{k}_i j \bm{k}_j}  \notag \\
        &- P_-(ij)  \sum_n \left( t_{i \bm{k}_i n \bm{k}_j}^{a \bm{k}_a b \bm{k}_b} \left( \tilde{\mathcal{F}}_{n \bm{k}_j j \bm{k}_j} +  \frac{1}{2} \sum_{c} t_{j \bm{k}_j}^{c \bm{k}_j} \tilde{\mathcal{F}}_{n \bm{k}_j}^{c \bm{k}_j} \right) \right)  \notag \\
        &- \sum_{c, d} \tilde{\eta}_{i j \bm{k}_i \bm{k}_j}^{c d} \bra{b \bm{k}_b a \bm{k}_a} \ket{c \bm{k}_i d \bm{k}_j}_{vvvv}
          + P_-(ab) \hat{\mathcal{G}}_{i \bm{k}_i j \bm{k}_j}^{a \bm{k}_a b \bm{k}_b}.
  \end{align}
  Intermediates $\tilde{\eta}_{i j \bm{k}_i \bm{k}_j}^{c d}$, $\hat{\mathcal{G}}_{i \bm{k}_i j \bm{k}_j}^{a \bm{k}_a b \bm{k}_b}$ and $\tilde{\mathcal{Y}}_{m \bm{k}_m n \bm{k}_n i \bm{k}_i j \bm{k}_j}$ are newly introduced.
  The definition of the intermediate $\tilde{\eta}_{i j \bm{k}_i \bm{k}_j}^{c d}$ is
  \begin{equation}
      \tilde{\eta}_{i j \bm{k}_i \bm{k}_j}^{c d} = t_{i \bm{k}_i}^{c \bm{k}_i} t_{j \bm{k}_j}^{d \bm{k}_j}.
  \end{equation}
  That of the intermediate $\hat{\mathcal{G}}_{i \bm{k}_i j \bm{k}_j}^{a \bm{k}_a b \bm{k}_b}$ is
  \begin{align}
      \hat{\mathcal{G}}_{i \bm{k}_i j \bm{k}_j}^{a \bm{k}_a b \bm{k}_b}
      =&\sum_r                                       \Biggl( \check{\rho}_r^b ( \bm{k}_b ) \biggl( \sum_{c, d} \tilde{\eta}_{i j \bm{k}_i \bm{k}_j}^{c d} \bra{r \bm{k}_b a \bm{k}_a} \ket{c \bm{k}_i d \bm{k}_j}_{ovvv}^{[ov]} \biggr. \Biggr.  \notag \\
        &~~~~~~~~~~~~~~~~~~~~~~~  \Biggl.                                                \biggl. + \left( \bra{i \bm{k}_i j \bm{k}_j}  \ket{r  \bm{k}_b a \bm{k}_a}_{ooov}^{[ov]} \right) ^* \biggr) \Biggr) .
  \end{align}
  That of the intermediate $\tilde{\mathcal{Y}}_{m \bm{k}_m n \bm{k}_n i \bm{k}_i j \bm{k}_j}$ is
  \begin{align}  \label{def_Y_oo}
       \tilde{\mathcal{Y}}_{m \bm{k}_m n \bm{k}_n i \bm{k}_i j \bm{k}_j}
       = \frac{1}{2} &           \biggl(    P_-(ij) \tilde{\mathcal{V}}_{m \bm{k}_m n \bm{k}_n i \bm{k}_i j \bm{k}_j}
                                                   + \bra{m \bm{k}_m n \bm{k}_n} \ket{i \bm{k}_i j \bm{k}_j}_{oooo}  \notag \\
                           & ~~~~  \biggl. + \sum_{c, d} \tilde{\zeta}_{ij}^{cd} ( \bm{k}_i, \bm{k}_j ) \bra{m \bm{k}_m n \bm{k}_n} \ket{c \bm{k}_i d \bm{k}_j}_{oovv}^{[vv]} \biggr) ,
  \end{align}
  where $\tilde{\mathcal{V}}_{m \bm{k}_m n \bm{k}_n i \bm{k}_i j \bm{k}_j}$is newly introduced intermediate defined as
  \begin{equation}
      \tilde{\mathcal{V}}_{m \bm{k}_m n \bm{k}_n i \bm{k}_i j \bm{k}_j} = \sum_c \hat{\rho}_j^c ( \bm{k}_j )  \bra{m \bm{k}_m n \bm{k}_n} \ket{i \bm{k}_i c \bm{k}_j}_{ooov}^{[ov]}.
  \end{equation}

 \section{Implementation of the presented method}  \label{ImplementMethod}
 
  In this section, a way of implementation of the presented method is described.
  In Section \ref{Sec_overall_flow}, overall flow of the presented implementation is described.
  In Section \ref{array_RHS}, arrays used for the right-hand sides of simultaneous equations are explained.
  In Section \ref{common_procedure}, procedures which are common to plural steps are described.
  In Section \ref{OurImplementaiton}, our implementation is given.

  \subsection{Overall flow}  \label{Sec_overall_flow}
  
  Overall flow of the presented method is as follows:
  
  \noindent
  The matrix elements of the Fock operator and antisymmetrized two-electron integrals are input as described in Section \ref{IO}.
  The initial values of single-excitation and double-excitation amplitudes are set as described in Section \ref{IO}.
  The right-hand sides of the simultaneous equations (\ref{EqSingle}) and (\ref{EqDouble}) are computed and single-excitation and double-excitation amplitudes are updated iteratively
  until these right-hand sides are sufficiently close to zero.
  Thus we obtain single-excitation and double-excitation amplitudes amplitudes that satisfy the CCSD equations as outputs.

  \subsection{Arrays for the right-hand sides of the simultaneous equations for single-excitation and double-excitation amplitudes}  \label{array_RHS}

  We prepare arrays for the right-hand sides of the simultaneous equations (\ref{EqSingle}) and (\ref{EqDouble}) for single-excitation and double-excitation amplitudes.
  Intermediate results during computation and the final results are stored in these arrays.
  Let us denote these arrays for single-excitation and double-excitation amplitudes by $S_	1$ and $S_2$, respectively.
  See Section \ref{RHS_abstract} for details.

  \subsection{Procedures which are common to plural steps in implementation}  \label{common_procedure}

  In the presented implementation, some procedures are common to plural steps.
  In this section, such procedures are described.

  \subsubsection{Procedures executed only in the diagonal processes}  \label{DiagProcedure}

  There are procedures executed only in the diagonal processes whose numbers are $\bar{\bm{k}} + \bar{\bm{k}} N_{\bm{k}}$ $( \bar{\bm{k}} = 0, 1, ..., N_{\bm{k}} - 1 )$.
  Let us call such a procedure ``Diagonal procedure''.

  \subsubsection{Summations which are concerned with the intermediates $\tilde{\mathcal{L}}$ and $\tilde{\mathcal{Z}}$}  \label{Sum_L_Z}
  
  In this section, we describe ways of computation of the summations
  $\sum_{\bm{k}_e} \tilde{\mathcal{Z}}_{p \bar{\bm{k}} q \bar{\bm{k}}}^{\bm{k}_e}$,
  $\sum_{\bm{k}_m} \tilde{\mathcal{Z}}_{\bm{k}_m}^{g \bar{\bm{k}} h \bar{\bm{k}}}$,
  $\sum_{\bm{k}_m} \tilde{\mathcal{L}}_{p \bar{\bm{k}} \bm{k}_m}^{g \bar{\bm{k}}}$ and
  $\sum_{\bm{k}_e} \tilde{\mathcal{L}}_{p \bar{\bm{k}}}^{g \bar{\bm{k}} \bm{k}_e}$
  in (\ref{EqSingle}), (\ref{def_F_oo}) and (\ref{def_F_vv}).
  The intermediates $\tilde{\mathcal{L}}_{p \bar{\bm{k}} \bm{k}_m}^{g \bar{\bm{k}}}$ and $\tilde{\mathcal{Z}}_{\bm{k}_m}^{g \bar{\bm{k}} h \bar{\bm{k}}}$
  are computed according to (\ref{def_Z_vv}) and (\ref{def_L_oo}), respectively,  in the process whose process number is $\bm{k}_m + \bar{\bm{k}} N_{\bm{k}}$.
  The intermediatess $\tilde{\mathcal{L}}_{p \bar{\bm{k}}}^{g \bar{\bm{k}} \bm{k}_e}$ and $\tilde{\mathcal{Z}}_{p \bar{\bm{k}} q \bar{\bm{k}}}^{\bm{k}_e}$
  are computed according to (\ref{def_Z_oo}) and (\ref{def_L_vv}), respectively,   in the process whose process number is $\bar{\bm{k}} + \bm{k}_e N_{\bm{k}}$.
  
\begin{figure}[H]
      \begin{center}
         \includegraphics[width=0.7\textwidth]{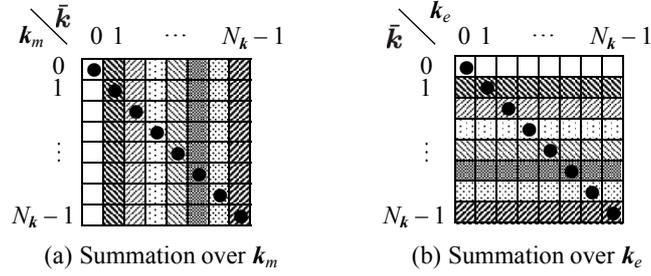}
         \caption{Summation over the indices $\bm{k}_m$ and $\bm{k}_e$}
         \label{Fig_sum_k}
      \end{center}
  \end{figure}

  The ways of the summations $\sum_{\bm{k}_m}$ and $\sum_{\bm{k}_e}$ are described using Fig. \ref{Fig_sum_k}.
  
  \noindent
  The summations over the indeices $\bm{k}_m$ and $\bm{k}_e$ are taken over processes in vertical and horizontal directions, respectively.
  The results are stored in the diagonal processes whose numbers are $\bar{\bm{k}} + \bar{\bm{k}} N_{\bm{k}}$ according to $\bar{\bm{k}}$.
  When we use MPI (Message Passing Interface), this procedure can be executed by the subroutine MPI\_REDUCE in a communicator grouping processes in the vertical or horizontal direction.

  \subsubsection{Change of an index to be identified through a process number}  \label{IndexChange}

\begin{figure}[H]
      \begin{center}
         \includegraphics[width=0.7\textwidth]{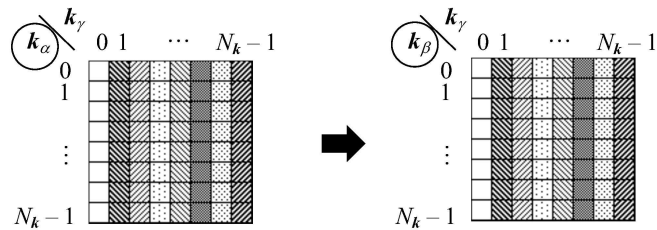}
         \caption{Change of index from $\bm{k}_{\alpha}$ to $\bm{k}_{\beta}$}
         \label{Fig_ExcRow}
      \end{center}
  \end{figure}

\begin{figure}[H]
      \begin{center}
         \includegraphics[width=0.7\textwidth]{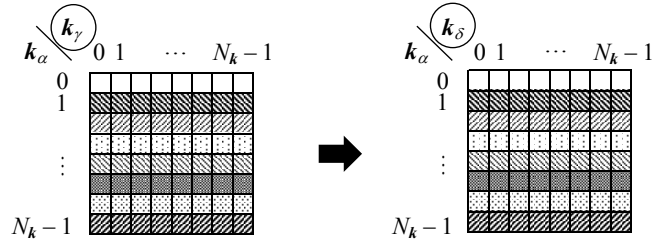}
         \caption{Change of index from $\bm{k}_{\gamma}$ to $\bm{k}_{\delta}$}
         \label{Fig_ExcCol}
      \end{center}
  \end{figure}

  Let us consider a status that quantities $\Theta_{\alpha \bm{k}_{\alpha} \beta \bm{k}_{\beta}}^{\gamma \bm{k}_{\gamma} \delta \bm{k}_{\delta}}$ are distributed to each process
  and two momentums $\bm{k}_{\alpha}$ and $\bm{k}_{\gamma}$ are identified through a process number by representing it as $\bm{k}_{\alpha} + \bm{k}_{\gamma} N_{\bm{k}}$.
  In the presented method, redistribution of these quantities to the following situations is necessary.
  \begin{itemize}
      \item Two momentums $\bm{k}_{\beta}$ and $\bm{k}_{\gamma}$ are identified through a process number by representing it as $\bm{k}_{\beta} + \bm{k}_{\gamma} N_{\bm{k}}$.
      \item Two momentums $\bm{k}_{\gamma}$ and $\bm{k}_{\delta}$ are identified through a process number by representing it as $\bm{k}_{\delta} + \bm{k}_{\gamma} N_{\bm{k}}$.
  \end{itemize}
  See Figs. \ref{Fig_ExcRow} and \ref{Fig_ExcCol}.
  When we use MPI, the first (second) change can be achieved by the subroutine MPI\_ALLTOALL in a communicator grouping processes in the vertical (horizontal) direction.
  
  \noindent

  \subsubsection{Summation between two quantities which have eight indices}  \label{Sum_88}
  
\begin{figure}[H]
      \begin{center}
         \includegraphics[width=0.7\textwidth]{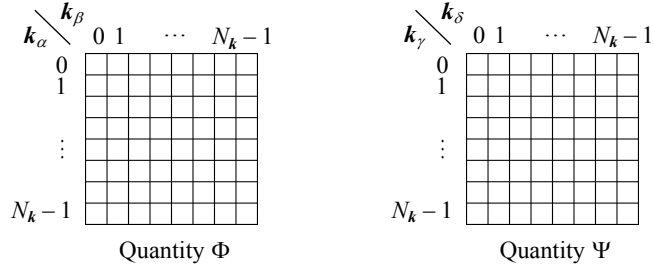}
         \caption{Processes for $\Phi$ and $\Psi$}
         \label{Fig_sum_88}
      \end{center}
  \end{figure}

  In this section, we consider a summation represented as
  \begin{equation}
      \Xi_{\alpha \bm{k}_{\alpha} \beta \bm{k}_{\beta} \gamma \bm{k}_{\gamma} \delta \bm{k}_{\delta}}
      = \sum_{\varepsilon , \bm{k}_{\varepsilon}, \zeta} \Phi_{\alpha \bm{k}_{\alpha} \beta \bm{k}_{\beta} \varepsilon \bm{k}_{\varepsilon} \zeta \bm{k}_{\zeta}}
                                                                                     \Psi_{\gamma \bm{k}_{\gamma} \delta \bm{k}_{\delta} \varepsilon \bm{k}_{\varepsilon} \zeta \bm{k}_{\zeta}}.
  \end{equation}
  In this section, indices do not represent kind of spin orbitals --- occupied or virtual.
  Assume that the elements of $\Phi$ and $\Psi$ are stored in processes whose numbers are $\bm{k}_{\alpha} + \bm{k}_{\beta} N_{\bm{k}}$ and $\bm{k}_{\gamma} + \bm{k}_{\delta} N_{\bm{k}}$, respectively,
  according to momentums $\bm{k}_{\alpha}$, $\bm{k}_{\beta}$, $\bm{k}_{\gamma}$ and $\bm{k}_{\delta}$.
  See Fig. \ref{Fig_sum_88}.
  
  \noindent
  To compute $\Xi$, it is necessary to send $\Psi$ to an appropriate process with consideration of conservation law of momentum.
  For each process as a receiver of $\Psi$, a process as a sender of $\Psi$ is uniquely determined when the momentum $\bm{k}_{\gamma}$ is specified because of conservation law of momentum.
  For $\Delta = 0, 1, ..., N_{\bm{k}} - 1$, we repeat the following procedure.
  For $\bm{k}_{\alpha}$ particular to a process, $\bm{k}_{\gamma}$ is specified by $\bm{k}_{\gamma} = \bm{k}_{\alpha} + \Delta \mod N_{\bm{k}}$.
  Then, $\bm{k}_{\delta}$ is uniquely determined from conservation law of momentum.
  The elements of $\Psi$ in the process whose process number is $\bm{k}_{\gamma} + \bm{k}_{\delta} N_{\bm{k}}$ are sent to the process whose process number is $\bm{k}_{\alpha} + \bm{k}_{\beta} N_{\bm{k}}$.
  When we use MPI, this procedure can be executed by the subroutines MPI\_SEND and MPI\_RECV.
  The elements of $\Xi$ are computed from $\Phi$ and the received $\Psi$.

  \subsubsection{Exchange of the first and the second indices identified through a process number}  \label{DiagSwap}
  
\begin{figure}[H]
      \begin{center}
         \includegraphics[width=0.7\textwidth]{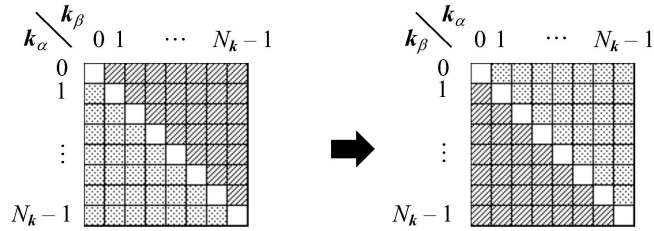}
         \caption{Change of a way of representing a process number}
         \label{Fig_diag_swap}
      \end{center}
  \end{figure}

  In our method, two momentums $\bm{k}_{\alpha}$ and $\bm{k}_{\beta}$ are identified through a process number by representing it as $\bm{k}_{\alpha} + \bm{k}_{\beta} N_{\bm{k}}$.
  In some steps in our method, to make an equivalent status under a condition that the two momentums are identified by representing a process number as $\bm{k}_{\beta} + \bm{k}_{\alpha} N_{\bm{k}}$ is necessary for particular quantities.
  See Fig. \ref{Fig_diag_swap}.
  This change can be achieved by transfer such quantities from a process whose process number is $\bm{k}_{\alpha} + \bm{k}_{\beta} N_{\bm{k}}$ to the one whose process number is $\bm{k}_{\beta} + \bm{k}_{\alpha} N_{\bm{k}}$ $( \alpha \neq \beta )$.
  Such quantities in the diagonal processes are copied to another array.
  When we use MPI, this procedure can be executed by the subroutines MPI\_SEND and MPI\_RECV.

  \subsection{Implementation}  \label{OurImplementaiton}

  In this section, we describe implementation of procedure after inputs and the initial values of single-excitation and double-excitation amplitudes are distributed to each process.
  This procedure is iterated until solution of the simultaneous equations (\ref{EqSingle}) and (\ref{EqDouble}) are obtained.

  \subsubsection{Computation of the intermediates $\hat{\rho}$, $\check{\rho}$ and $\tilde{\zeta}$}

  Let us consider a case such that a process number is represented as $\bm{k}_{\alpha} + \bm{k}_{\beta} N_{\bm{k}}$.
  From single-excitation amplitudes, the intermediates $\hat{\rho}$ and $\check{\rho}$ are set according to (\ref{def_hat_rho}) and (\ref{def_check_rho}),
  and the intermediate $\tilde{\zeta}$ is computed through (\ref{def_zeta}).

  \subsubsection{Computation of the intermediates $\tilde{\mathcal{F}}$}
  
  The summations $\sum_{\bm{k}_e} \tilde{\mathcal{Z}}_{p \bar{\bm{k}} q \bar{\bm{k}}}^{\bm{k}_e}$,
  and $\sum_{\bm{k}_m} \tilde{\mathcal{Z}}_{\bm{k}_m}^{g \bar{\bm{k}} h \bar{\bm{k}}}$ in (\ref{def_F_oo}) and in (\ref{def_F_vv})
  are obtained from the way described in Section \ref{Sum_L_Z} and are stored in the diagonal processes.

  The intermediate $\tilde{\mathcal{K}}$ is computed according to (\ref{def_K}).
  It is Diagonal procedure described in Section \ref{DiagProcedure}.

  The intermediates $\tilde{\mathcal{F}}$ are computed according to (\ref{def_F_oo}), (\ref{def_F_vv}) and (\ref{def_F_ov}).
  It is Diagonal procedure described in Section \ref{DiagProcedure}.

\begin{figure}[H]
      \begin{center}
         \includegraphics[width=0.7\textwidth]{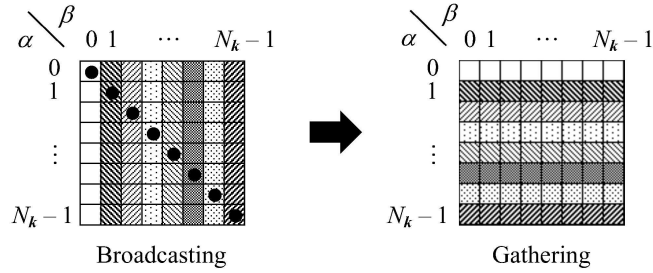}
         \caption{Broadcasting and gathering of the intermediates $\tilde{\mathcal{F}}$}
         \label{Fig_share_F}
      \end{center}
  \end{figure}

  Elements of the intermediates $\tilde{\mathcal{F}}$ are separately stored in the diagonal processes according to $\bar{\bm{k}}$ at this stage.
  It is necessary that each process stores all the elements of $\tilde{\mathcal{F}}$.
  We describe a procedure using Fig. \ref{Fig_share_F}.
  The elements of $\tilde{\mathcal{F}}$ in the diagonal processes are broadcasted to the other processes in the vertical direction.
  When we use MPI, this procedure can be executed by the subroutine MPI\_BCAST in a communicator grouping processes in the vertical direction.
  After broadcasting, these results are gathered to each process in the horizontal direction.
  When we use MPI, this procedure can be executed by the subroutine MPI\_ALLGATHER in a communicator grouping processes in the horizontal direction.
  Thus, all the intermediates of $\tilde{\mathcal{F}}$ are stored in all the processes.

  \subsubsection{Computation of the right-hand sides of simultaneous equations for single-excitation amplitudes}

  We compute the right-hand side of (\ref{EqSingle}).
  The first item can directly be substituted to the array $S_1$.
  This is Diagonal procedure described in Section \ref{DiagProcedure}.
  On the other items, when computation of each item is finished, addition to or subtraction from the array $S_1$ is done.
  This is also Diagonal procedure.
  Computation of from the second to the fifth items is Diagonal procedure and is straightforward.
  Computation of the sixth and the seventh items is as described in Section \ref{Sum_L_Z}.
  
  Thus, the right-hand sides of simultaneous equations for single-excitation amplitudes are obtained in the diagonal processes.

  \subsubsection{Computation of the intermediate $\tilde{\mathcal{W}}$}
  
  The intermediate $\tilde{\mathcal{W}}$ is computed according to (\ref{rearranged_W}).
  
  We describe computation of the first term in the right-hand side.
  At the beginning of this computation, the indices $\bm{k}_j$ and $\bm{k}_x$ of double-excitation amplitudes $t_{j \bm{k}_j y \bm{k}_y}^{x \bm{k}_x b \bm{k}_b}$
  are identified through a process number by representing it as $\bm{k}_j + \bm{k}_x N_{\bm{k}}$.
  It is desired that the indices $\bm{k}_j$ and $\bm{k}_b$ of double-excitation amplitudes are identified through a process number by representing it as $\bm{k}_j + \bm{k}_b N_{\bm{k}}$.
  For this purpose, the elements of double-excitation amplitudes are exchanged between processes by the procedure described in Section \ref{IndexChange}.
  See also Fig. \ref{Fig_ExcCol}.
  After this exchange, the term is obtained using the procedure described in Section \ref{Sum_88}.
  
  Addition of the second term in the right-hand side is straightforward.

  On the third term in the right-hand side, note that antisymmetrized two-electron integrals used in computation of the intermediate $\tilde{\mathcal{J}}$ is stored in a process whose process number is $\bm{k}_b + \bm{k}_j N_{\bm{k}}$.
  The elements $\tilde{\mathcal{J}}_{j \bm{k}_j n \bm{k}_n}^{b \bm{k}_b f \bm{k}_f}$ are computed according to (\ref{def_J}) in a process whose process number is $\bm{k}_b + \bm{k}_j N_{\bm{k}}$.
  After computation, the procedure described in Section \ref{DiagSwap} is executed.
  Then, the third term is obtained in a process whose process number is $\bm{k}_j + \bm{k}_b N_{\bm{k}}$.
  This term is added.
  
  Thus, the intermediate $\tilde{\mathcal{W}}$ is obtained.

  \subsubsection{Computation of the intermediate $\tilde{\mathcal{C}}$ and operation by the operators $P_-(ij) P_-(ab)$ to it}  \label{Compute_PPC}
  
  The intermediate $\tilde{\mathcal{C}}$ is computed according to (\ref{def_C}).
  The first term in the right-hand side is computed by the procedure described in Section \ref{Sum_88}.
  The result is substituted to the array $S_2$.
  After computation of the second term, the result is subtracted from the array $S_2$.
  At this stage, it holds $S_2 = \tilde{\mathcal{C}}_{i \bm{k}_i j \bm{k}_j}^{a \bm{k}_a b \bm{k}_b}$.
  
  The result of operation by the operators $P_-(ij) P_-(ab)$ is concretely
  \begin{equation}  \label{concrete_PPC}
      P_-(ij) P_-(ab) \tilde{\mathcal{C}}_{i \bm{k}_i j \bm{k}_j}^{a \bm{k}_a b \bm{k}_b}
      =  \tilde{\mathcal{C}}_{i \bm{k}_i j \bm{k}_j}^{a \bm{k}_a b \bm{k}_b}
        - \tilde{\mathcal{C}}_{i \bm{k}_i j \bm{k}_j}^{b \bm{k}_b a \bm{k}_a}
        - \tilde{\mathcal{C}}_{j \bm{k}_j i \bm{k}_i}^{a \bm{k}_a b \bm{k}_b}
        + \tilde{\mathcal{C}}_{j \bm{k}_j i \bm{k}_i}^{b \bm{k}_b a \bm{k}_a}.
  \end{equation}
  When computation of $\tilde{\mathcal{C}}$ is finished, the elements of the first term in the right-hand side are distributed to each process according to the indices $\bm{k}_i$ and $\bm{k}_a$
  by representing process numbers as $\bm{k}_i + \bm{k}_a N_{\bm{k}}$.
  In computation of the other items, redistribution of the elements of $\tilde{\mathcal{C}}$ is necessary and we prepare another array different from $S_2$ for this purpose.
  The array $S_2$ is copied to this array before redistribution.
  To subtract the second (third) term, the elements of $\tilde{\mathcal{C}}$ should be redistributed according to the indices $\bm{k}_i$ and $\bm{k}_b$ ($\bm{k}_j$ and $\bm{k}_a$)
  by representing process numbers as $\bm{k}_i + \bm{k}_b N_{\bm{k}}$ ($\bm{k}_j + \bm{k}_a N_{\bm{k}}$).
  This redistribution of the elements can be achieved by the procedure described in Section \ref{IndexChange}.
  To add the fourth term, the elements of $\tilde{\mathcal{C}}$ should be redistributed according to the indices $\bm{k}_j$ and $\bm{k}_b$ by representing process numbers as $\bm{k}_j + \bm{k}_b N_{\bm{k}}$.
  This redistribution of the elements can be achieved by the procedure described in Section \ref{IndexChange} starting from the redistributed configuration of the elements of $\tilde{\mathcal{C}}$ for the subtraction in the second term.
  After each redistribution is finished, addition to or subtraction from the array $S_2$ is done.

  \subsubsection{Computation of the term $\bar{\mathcal{A}}$}
  
  The term $\bar{\mathcal{A}}$ in (\ref{EqDouble}) is computed according to he right-hand side of (\ref{A_1st}).
  The first term is obtained by the procedure described in Section \ref{Compute_PPC} and the result is stored in the array $S_2$.
  Addition of the second term to the array $S_2$ is straightforward.
  Note that the third term is computed in processes whose process numbers are represented not as $\bm{k}_i+ \bm{k}_a N_{\bm{k}}$ but as $\bm{k}_a+ \bm{k}_i N_{\bm{k}}$.
  After finish of this computation, the procedure described in Section \ref{DiagSwap} is executed to transfer the results to the appropriate processes.
  Then, the transferred results are added to the array $S_2$.
  At this stage, it holds $S_2 = \bar{\mathcal{A}}$.

  \subsubsection{Computation of the term $\check{\mathcal{A}}$ -- Before redistribution of double-excitation amplitudes}
  
  The elements of the term $\check{\mathcal{A}}$ are computed in processes whose process numbers are $\bm{k}_i+ \bm{k}_j N_{\bm{k}}$.
  Intermediate results during computation and the final results are stored in an array $\check{S}_2$ which is different from $S_2$.
  Double-excitation amplitudes are redistributed according to the indices $\bm{k}_i$ and $\bm{k}_j$ by representing process numbers as $\bm{k}_i + \bm{k}_j N_{\bm{k}}$.
  The first and the second term in (\ref{A_2nd}) are computed before this redistribution.
  The first term and the intermediate $\check{\mathcal{G}}$ are computed.
  After this computation, it holds
  \begin{equation}  \label{S2_intermediate}
      \check{S}_2 =  \sum_{r, s} \tilde{\eta}_{r s}^{a b \bm{k}_a \bm{k}_b} \bra{r \bm{k}_a s \bm{k}_b} \ket{i \bm{k}_i j \bm{k}_j}_{oooo} + \check{\mathcal{G}}_{i \bm{k}_i j \bm{k}_j}^{a \bm{k}_a b \bm{k}_b}.
  \end{equation}
  and the intermediate $\check{\mathcal{G}}$ is stored in another array.
  The procedure described in Section \ref{DiagSwap} for operation of the operator $P_-(ij)$ is applied to the intermediate $\check{\mathcal{G}}$.
  The elements $\check{\mathcal{G}}_{j \bm{k}_j i \bm{k}_i}^{a \bm{k}_a b \bm{k}_b}$ stored in the processes whose process numbers are $\bm{k}_j + \bm{k}_i N_{\bm{k}}$ are subtracted from the array $\check{S}_2$ in (\ref{S2_intermediate}).
  After this procedure, it holds
  \begin{equation}
      \check{S}_2 =  \sum_{r, s} \tilde{\eta}_{r s}^{a b \bm{k}_a \bm{k}_b} \bra{r \bm{k}_a s \bm{k}_b} \ket{i \bm{k}_i j \bm{k}_j}_{oooo} + P_-(ij) \check{\mathcal{G}}_{i \bm{k}_i j \bm{k}_j}^{a \bm{k}_a b \bm{k}_b}.
  \end{equation}

  \subsubsection{Redistribution of double-excitation amplitudes}  \label{Redist_1st}

  At the beginning of computation of $\check{\mathcal{A}}$, double-excitation amplitudes $t_{i \bm{k}_i j \bm{k}_j}^{a \bm{k}_a b \bm{k}_b}$
  are distributed according to the indices $\bm{k}_i$ and $\bm{k}_a$ by representing process numbers as $\bm{k}_i + \bm{k}_a N_{\bm{k}}$.
  The amplitudes should be redistributed according to the indices $\bm{k}_i$ and $\bm{k}_j$  by representing process numbers as $\bm{k}_i + \bm{k}_j N_{\bm{k}}$.
  For this purpose, elements of double-excitation amplitudes are exchanged between processes by the procedure described in Section \ref{IndexChange}.
  See also Fig. \ref{Fig_ExcCol}.
  The original array in which the amplitudes are stored should be kept.

  \subsubsection{Computation of the term $\check{\mathcal{A}}$ -- After redistribution of double-excitation amplitudes}
  
  In computation of the third item in the right-hand side of (\ref{A_2nd}), no communication between processes occurs.
  When this computation is finished, it holds
  \begin{equation}
      \check{S}_2 = \check{\mathcal{A}} - \sum_{e, \bm{k}_e, f} t_{i \bm{k}_i j \bm{k}_j}^{e \bm{k}_e f \bm{k}_f} \tilde{\mathcal{Y}}^{a \bm{k}_a b \bm{k}_b e \bm{k}_e f \bm{k}_f}.
  \end{equation}

  Next, we consider computation of the fourth item in the right-hand side of (\ref{A_2nd}).
  Note that four arithmetic operations in computation of the intermediate $\tilde{\mathcal{Y}}$ given in (\ref{def_Y_vv}) are done in processes whose process numbers are $\bm{k}_b + \bm{k}_a N_{\bm{k}}$.
  After computation of the intermediate $\tilde{\mathcal{V}}$,
  its elements are copied to the process whose process number is $\bm{k}_a + \bm{k}_b N_{\bm{k}}$ through the procedure described in Section \ref{DiagSwap}.
  The copied elements are subtracted from the ones of the original $\tilde{\mathcal{V}}$.
  Thus, computation of the item $P_-(ab) \tilde{\mathcal{V}}$ is finished.
  After this step, the elements of the intermediate $\tilde{\mathcal{Y}}$ are obtained according to (\ref{def_Y_vv}) without communication between processes.
  The summation $\sum_{e, \bm{k}_e, f}$ is executed through the procedure described in Section \ref{Sum_88}.
  The results are added to the array $\check{S}_2$.
  Now, it holds $\check{S}_2 = \check{\mathcal{A}}$.

  \subsubsection{Addition of the term $\check{\mathcal{A}}$ to the array $S_2$}
  
  The elements of the term $\check{\mathcal{A}}_{i \bm{k}_i j \bm{k}_j}^{a \bm{k}_a b \bm{k}_b}$ are stored in processes whose numbers are $\bm{k}_i + \bm{k}_j N_{\bm{k}}$.
  To add them to the array $S_2$, redistribution of them is necessary.
  The elements should be redistributed according to the indices $\bm{k}_i$ and $\bm{k}_a$ to processes whose numbers are $\bm{k}_i + \bm{k}_a N_{\bm{k}}$.
  For this purpose, the elements of the array $\check{S}_2$ are exchanged between processes by the procedure described in Section \ref{IndexChange}.
  See also Fig. \ref{Fig_ExcCol}.
  The redistributed elements are added to the array $S_2$.
  Now, it holds $S_2 = \bar{\mathcal{A}} + \check{\mathcal{A}}$.

  \subsubsection{Redistribution of double-excitation amplitudes}

  We redistribute double-excitation amplitudes $t_{i \bm{k}_i j \bm{k}_j}^{a \bm{k}_a b \bm{k}_b}$ again for computation of the terms $\tilde{\mathcal{A}}$ and $\hat{\mathcal{A}}$.
  These amplitudes are originally distributed according to the indices $\bm{k}_i$ and $\bm{k}_a$ by representing process numbers as $\bm{k}_i + \bm{k}_a N_{\bm{k}}$.
  The amplitudes should be redistributed according to the indices $\bm{k}_b$ and $\bm{k}_a$  by representing process numbers as $\bm{k}_b + \bm{k}_a N_{\bm{k}}$.
  For this purpose, elements of double-excitation amplitudes are exchanged between processes by the procedure described in Section \ref{IndexChange}.
  See also Fig. \ref{Fig_ExcRow}.
  The amplitudes which have been distributed in Section \ref{Redist_1st} for computation of $\check{\mathcal{A}}$ should be kept for computation of the intermediate $\tilde{\mathcal{X}}$ defined in (\ref{def_X}).

  \subsubsection{Computation of the intermediate $\tilde{\mathcal{X}}$}

  The intermediate $\tilde{\mathcal{X}}_{m \bm{k}_m n \bm{k}_n i \bm{k}_i j \bm{k}_j}$ defined in (\ref{def_X}) is obtained using the procedure described in Section \ref{Sum_88}.
  The elements of $\tilde{\mathcal{X}}$ are distributed according to the indices $\bm{k}_i$ and $\bm{k}_j$ by representing process numbers as $\bm{k}_i + \bm{k}_j N_{\bm{k}}$.

  \subsubsection{Computation of the term $\tilde{\mathcal{A}}$}

  The elements of the terms $\tilde{\mathcal{A}}_{i \bm{k}_i j \bm{k}_j}^{a \bm{k}_a b \bm{k}_b}$ and $\hat{\mathcal{A}}_{i \bm{k}_i j \bm{k}_j}^{a \bm{k}_a b \bm{k}_b}$
  are computed in processes whose process numbers are $\bm{k}_b + \bm{k}_a N_{\bm{k}}$.
  Intermediate results during computation and the final results of  $\tilde{\mathcal{A}} + \hat{\mathcal{A}}$ are stored in an array $\hat{S}_2$ which is different from $S_2$ and $\check{S}_2$.

  The term $\tilde{\mathcal{A}}$ defined in (\ref{A_3rd}) can be obtained using the procedure described in Section \ref{Sum_88}.
  The results are substituted to the array $\hat{S}_2$.
  Then, it holds $\hat{S}_2 = \tilde{\mathcal{A}}$.

  \subsubsection{Computation of the term $\hat{\mathcal{A}}$ -- The first term}
  
  We consider computation of the first item in the right-hand side of (\ref{A_4th}).
  Note that four arithmetic operations in computation of the intermediate $\tilde{\mathcal{Y}}$ given in (\ref{def_Y_oo}) are done in processes whose process numbers are $\bm{k}_i + \bm{k}_j N_{\bm{k}}$.
  After computation of the intermediate $\tilde{\mathcal{V}}$,
  its elements are copied to the process whose process number is $\bm{k}_j + \bm{k}_i N_{\bm{k}}$ through the procedure described in Section \ref{DiagSwap}.
  The copied elements are subtracted from the ones of the original $\tilde{\mathcal{V}}$.
  Thus, computation of the item $P_-(ij) \tilde{\mathcal{V}}$ is finished.
  After this step, the elements of the intermediate $\tilde{\mathcal{Y}}$ are obtained according to (\ref{def_Y_oo}) without communication between processes.
  The summation $\sum_{m, \bm{k}_m, n}$ is executed through the procedure described in Section \ref{Sum_88}.
  The results are added to the array $\hat{S}_2$.
  Now, it holds $\hat{S}_2 = \tilde{\mathcal{A}} + \sum_{m, \bm{k}_m, n} t_{m \bm{k}_m n \bm{k}_n}^{a \bm{k}_a b \bm{k}_b} \tilde{\mathcal{Y}}_{m \bm{k}_m n \bm{k}_n i \bm{k}_i j \bm{k}_j}$.

  \subsubsection{Computation of the term $\hat{\mathcal{A}}$ -- The second term}
  
  In computation of the third item in the right-hand side of (\ref{A_4th}), no communication between processes occurs.
  When this computation is finished, it holds
  \begin{align}
      \hat{S}_2 =&   \tilde{\mathcal{A}}
                          + \sum_{m, \bm{k}_m, n} t_{m \bm{k}_m n \bm{k}_n}^{a \bm{k}_a b \bm{k}_b} \tilde{\mathcal{Y}}_{m \bm{k}_m n \bm{k}_n i \bm{k}_i j \bm{k}_j}  \notag  \\
                        &- P_-(ij)  \sum_n \left( t_{i \bm{k}_i n \bm{k}_j}^{a \bm{k}_a b \bm{k}_b} \left( \tilde{\mathcal{F}}_{n \bm{k}_j j \bm{k}_j} +  \frac{1}{2} \sum_{c} t_{j \bm{k}_j}^{c \bm{k}_j} \tilde{\mathcal{F}}_{n \bm{k}_j}^{c \bm{k}_j} \right) \right) .
  \end{align}

  \subsubsection{Computation of the term $\hat{\mathcal{A}}$ -- The third and the fourth terms}
  
  The third term and the intermediate $\hat{\mathcal{G}}$ are computed.
  The results are subtracted from or added to $\hat{S}_2$ and the intermediate $\hat{\mathcal{G}}$ is stored in another array.
  After this procedure, it holds
  \begin{equation}
      \hat{S}_2 = \tilde{\mathcal{A}}_{i \bm{k}_i j \bm{k}_j}^{a \bm{k}_a b \bm{k}_b} + \hat{\mathcal{A}}_{i \bm{k}_i j \bm{k}_j}^{a \bm{k}_a b \bm{k}_b} + \check{\mathcal{G}}_{i \bm{k}_i j \bm{k}_j}^{b \bm{k}_b a \bm{k}_a}.
  \end{equation}
  The procedure described in Section \ref{DiagSwap} for operation of the operator $P_-(ab)$.
  The elements $\check{\mathcal{G}}_{i \bm{k}_i j \bm{k}_j}^{b \bm{k}_b a \bm{k}_a}$ stored in the processes whose process numbers are $\bm{k}_a + \bm{k}_b N_{\bm{k}}$ are subtracted from the array $\hat{S}_2$.
  After this procedure, it holds $\hat{S}_2 = \tilde{\mathcal{A}} + \hat{\mathcal{A}}$.

  \subsubsection{Addition of the term $( \tilde{\mathcal{A}} + \hat{\mathcal{A}})$ to the array $S_2$}
  
  The elements of the term $( \tilde{\mathcal{A}}_{i \bm{k}_i j \bm{k}_j}^{a \bm{k}_a b \bm{k}_b} + \hat{\mathcal{A}}_{i \bm{k}_i j \bm{k}_j}^{a \bm{k}_a b \bm{k}_b} )$ are stored in processes whose numbers are $\bm{k}_b + \bm{k}_a N_{\bm{k}}$.
  To add them to the array $S_2$, redistribution of them is necessary.
  The elements should be redistributed according to the indices $\bm{k}_i$ and $\bm{k}_a$ to processes whose numbers are $\bm{k}_i + \bm{k}_a N_{\bm{k}}$.
  For this purpose, the elements of the array $\hat{S}_2$ are exchanged between processes by the procedure described in Section \ref{IndexChange}.
  See also Fig. \ref{Fig_ExcRow}.
  The redistributed elements are added to the array $S_2$.
  Now, it holds $S_2 = \bar{\mathcal{A}} + \check{\mathcal{A}} + \tilde{\mathcal{A}} + \hat{\mathcal{A}}$.

  \subsubsection{Convergence judgment to the solutions}
  
  Since the correct single-excitation and double-excitation amplitudes satisfy (\ref{criterion_single}) and (\ref{criterion_double}), respectively, the simplest convergence criterion for these amplitudes should be the one such that
  \begin{gather}
      \left| ( R_1 )_{g \bar{\bm{k}}}^{p \bar{\bm{k}}} \right| < \varepsilon_1, \label{criterion_single}  \\
      \left| ( R_2 )_{i {\bm{k}}_i j {\bm{k}}_j}^{a {\bm{k}}_a b {\bm{k}}_b} \right| < \varepsilon_2,  \label{criterion_double}
  \end{gather}
  where $\varepsilon_1$ and $\varepsilon_2$ are some constants, for all the indices.
  We continue to update both of single-excitation and double-excitation amplitudes via successive substitution until this criterion is satisfied.

  \subsubsection{Update of single-excitation and double-excitation amplitudes}
  
  We introduce the following quantities defined as
  \begin{gather}
      D_{i \bar{\bm{k}}}^{a \bar{\bm{k}}} = f_{i \bar{\bm{k}} i \bar{\bm{k}}} - f^{a \bar{\bm{k}} a \bar{\bm{k}}}, \\
      D_{i \bm{k}_i j \bm{k}_j}^{a \bm{k}_a b \bm{k}_b} = f_{i \bm{k}_i i \bm{k}_i} + f_{j \bm{k}_j j \bm{k}_j} - f^{a \bm{k}_a a \bm{k}_a} - f^{b \bm{k}_b b \bm{k}_b}.
  \end{gather}
  Let us denote the updated single-excitation and double-excitation amplitudes by $\breve{t}_{i \bar{\bm{k}}}^{a \bar{\bm{k}}}$ and $\breve{t}_{i \bm{k}_i j \bm{k}_j}^{a \bm{k}_a b \bm{k}_b}$, respectively.
  Then, we update the amplitudes as
  \begin{gather}
      \breve{t}_{i \bar{\bm{k}}}^{a \bar{\bm{k}}} = t_{i \bar{\bm{k}}}^{a \bar{\bm{k}}} + \left( D_{i \bar{\bm{k}}}^{a \bar{\bm{k}}} \right) ^{-1} ( R_1 )_{g \bar{\bm{k}}}^{p \bar{\bm{k}}}, \\
      \breve{t}_{i \bm{k}_i j \bm{k}_j}^{a \bm{k}_a b \bm{k}_b} =    t_{i \bm{k}_i j \bm{k}_j}^{a \bm{k}_a b \bm{k}_b}
                                                                                                  + \left( D_{i \bm{k}_i j \bm{k}_j}^{a \bm{k}_a b \bm{k}_b} \right) ^{-1} ( R_2 )_{i \bm{k}_i j \bm{k}_j}^{a \bm{k}_a b \bm{k}_b}.
  \end{gather}
  They are used in the next iteration.

 \section{Conclusion}  \label{Conclusion}
 
  A parallel computing method for the Coupled-Cluster Singles and Doubles (CCSD) in periodic systems is presented.
  The presented method uses $N_{\bm{k}}^2$ for the number of $k$ points since two indices which represent momentum attached to quantities which have eight indices are identified through a process number in parallel computing.
  The orders of computational cost and required memory space in each process are reduced by $N_{\bm{k}}^2$ compared with a sequential method.
  In implementation of the presented method, communication between processes in parallel computing appears in the outmost loop in a nested loop but does not appear inner nested loop.

  \section*{Acknowledgments}
 
  This research was supported by MEXT as Exploratory Challenge on Post-K computer'' (Frontiers of Basic Science: Challenging the Limits, Challenge of Basic Science: Fundamental Quantum Mechanics and Informatics).


\begin{thebibliography}{99}
    \bibitem{HK64}  P. Hohenberg and W. Kohn,
                              Phys. Rev, {\bf 136}, B864 (1964)
    \bibitem{KS65}  W. Kohn and L. J. Sham,
                              Phys. Rev., {\bf 140}, A1133 (1965)
    \bibitem{White92}  S. R. White,
                              Phys. Rev. Lett., {\bf 69}, 2863 (1992)
    \bibitem{WM99}  S. R. White and R. L. Martin,
                              The Journal of Chemical Physics, {\bf 110}, 4127 (1999)
    \bibitem{BH69}  S. F. Boys and N. C. Handy,
                              Proceedings of the Royal Society of London A: Mathematical, Physical and Engineering Sciences, {\bf 309}, 209 (1969),
                              http://rspa.royalsocietypublishing.org/content/309/1497/209.full.pdf
    \bibitem{UTOSC05}  N. Umezawa, S. Tsuneyuki, T. Ohno, K. Shiraishi and T. Chikyow,
                              The Journal of Chemical Physics, {\bf 122}, 224101 (2005)
    \bibitem{OAT17}  M. Ochi, R. Arita and S. Tsuneyuki,
                              Phys. Rev. Lett., {\bf 118}, 026402 (2017)
    \bibitem{Potthoff03}  M. Potthoff,
                              Eur. Phys. J. B, {\bf 32}, 429 (2003)
    \bibitem{Potthoff14}  M. Potthoff,
                              arXiv:1407.4065 [cond-mat.str-el] (2014)
    \bibitem{KNFM18}  T. Kosugi, H. Nishi, Y. Furukawa, and Y.-i. Matsushita,
                              The Journal of chemical physics, {\bf 148}, 224103 (2018)
    \bibitem{HJO00}  T. Helgaker, P. J{\o}rgensen and J. Olsen,
                              \underline{Molecular Electronic-Structure} \underline{Theory} (Wiley, 2000)
    \bibitem{MLWMRLBC16}  J. McClain, J. Lischner, T. Watson, D. A. Matthews, E. Ronca, S. G. Louie, T. C. Berkelbach, and G. K.-L. Chan,
                              Phys. Rev. B, {\bf 93}, 235139 (2016)
    \bibitem{MSCB17}  J. McClain, Q. Sun, G. K.-L. Chan, and Timothy C. Berkelbach,
                              Journal of Chemical Theory and Computation, {\bf 13}, 1209 (2017)
    \bibitem{FKNM18}  Y. Furukawa, T. Kosugi, H. Nishi and Y.-i. Matsushita,
                              The Journal of Chemical Physics, {\bf 148}, 204109 (2018)
    \bibitem{NKFM18}  H. Nishi, T. Kosugi, Y. Furukawa and Y.-i. Matsushita,
                              The Journal of Chemical Physics, {\bf 149}, 034106 (2018)
    \bibitem{PK18}  B. Peng and K. Kowalski,
                             Journal of Chemical Theory and Computation, {\bf 149}, 4335 (2018)
    \bibitem{Hedin65}  L. Hedin,
                             Phys. Rev., {\bf 139}, A796 (1965)
    \bibitem{LB18}  M. F. Lange, and Timothy C. Berkelbach,
                             Journal of Chemical Theory and Computation, {\bf 14}, 4224 (2018)
    \bibitem{KM19}  T. Kosugi and Y.-i. Matsushita,
                             The Journal of chemical physics, {\bf 150}, 114104 (2019)
    \bibitem{YS19}  T. Yamashita and T. Sakurai,
                             in preparation
    \bibitem{XCQZYX12}  Z. Y. Xie, J. Chen, M. P. Qin, J. W. Zhu, L. P. Yang and T. Xiang,
                             Phys. Rev. B, {\bf 86}, 045319 (2012)
    \bibitem{AKYY19}  S. Akiyama, Y. Kuramashi, T. Yamashita and Y. Yoshimura,
                             Phys. Rev. D, {\bf 100}, 054510 (2019)
    \bibitem{GS95}  J. Gauss and J. F. Stanton,
                              The Journal of Chemical Physics, {\bf 103}, 3561 (1995)
    \bibitem{HPTB04}  S. Hirata, R. Podeszwa, M. Tobita and R. J. Bartlett,
                              The Journal of Chemical Physics, {\bf 120}, 2581 (2004)
  \end{thebibliography}
\end{document}